\documentclass{vldb}

\usepackage{amsmath}

\usepackage{amssymb}

\usepackage{faktor}

\usepackage[inference]{semantic}

\usepackage{fancyvrb}

\usepackage{graphicx}

\usepackage{subfig}

\usepackage{float}

\usepackage{caption}

\usepackage{mdframed}

\usepackage{multirow}

\usepackage{balance}

\usepackage{fixltx2e}

\usepackage{setspace}
\usepackage[scaled]{beramono}
\usepackage[T1]{fontenc}

\usepackage{optional}

\usepackage{xstring}

\usepackage[table]{xcolor}

\usepackage{soul}

\usepackage[normalem]{ulem}

\usepackage{hyperref}

\usepackage[all]{hypcap}

\newenvironment{compact_enum}
{\setlength{\leftmargini}{1em}
\begin{enumerate}
  \setlength{\labelsep}{.3em} 
  \setlength{\itemsep}{.4em}
  \setlength{\parskip}{0pt}
  \setlength{\parsep}{0pt}}
{\end{enumerate}}

\def\semijoin{\mbox{$\mathrel{\raise1pt\hbox{\vrule height5pt depth0pt\hskip-1.5pt$>$\hskip -2.5pt$<$}}$}}

\newcommand{\gn}[1]  {\textit{#1}}           %
\newcommand{\gl}[1]  {\texttt{\textbf{#1}}}  %
\newcommand{\gs}[1]  {\textit{#1}}           %
\newcommand{\gp}[0]  {$\rightarrow$}         %
\newcommand{\gd}[0]  {$|$}                   %

\chardef\singlequote=13

\DeclareTextCommand{\_}{T1}{\leavevmode \kern.06em\vbox{\hrule width.4em}}

\usepackage{adjustbox}
\newcolumntype{R}[2]{%
    >{\adjustbox{angle=#1,lap=\width-(#2)}\bgroup}%
    l%
    <{\egroup}%
}

\newcolumntype{B}{|@{~}r@{~}|@{~}l@{~}c@{~}l@{~}|}  %
\newcolumntype{E}{@{}l@{~}c@{~}l@{}}                %
\newcolumntype{N}{@{}l@{}}                          %

\def\env{\mathrm{\Gamma}}

\usepackage{pifont}
\usepackage{wasysym}

\def\yes{$\checkmark$}
\def\error{$\times$}
\def\true{\texttt{t}}
\def\false{\texttt{f}}
\def\nulloption{\texttt{n}}
\def\missing{\texttt{m}}
\def\counter{\texttt{c}}
\def\partial{\LEFTcircle}
\def\irrelevant{-}
\def\inconsistent{\ding{106}}

\newcommand{\highlight}[1]{\noindent\textbf{#1:}}

\begin{document}

\title{The SQL++ Query Language: \\ Configurable, Unifying and Semi-structured%
\titlenote{This work was supported by NSF DC 0910820, NSF III 1018961, NSF IIS 1237174 and Informatica grants. The grants' PI was Prof Papakonstantinou who is a shareholder of an entity that commercializes outcomes of this research.
}}

\numberofauthors{1}
\author{
\alignauthor
Kian Win Ong, Yannis Papakonstantinou, Romain Vernoux\\
\{kianwin,yannis,rvernoux\}@cs.ucsd.edu
}

\maketitle

\begin{sloppypar}

\begin{abstract}

NoSQL databases support semi-structured data, typically modeled as JSON. They also provide limited (but expanding) query languages. Their idiomatic, non-SQL language constructs, the many variations, and the lack of formal semantics inhibit deep understanding of the query languages, and also impede progress towards clean, powerful, declarative query languages. 

This paper specifies the syntax and semantics of SQL++, which is applicable to both JSON native stores and SQL databases. The SQL++ semi-structured data model is a superset of both JSON and the SQL data model. SQL++ offers powerful computational capabilities for processing semi-structured data akin to prior non-relational query languages, notably OQL and XQuery. Yet, SQL++ is SQL backwards compatible and is generalized towards JSON by introducing only a small number of query language extensions to SQL. Indeed, the SQL capabilities are most often extended by removing semantic restrictions of SQL, rather than inventing new features. 

Recognizing that a query language standard is probably premature for the fast evolving area of NoSQL databases, SQL++ includes configuration options that formally itemize the semantics variations that language designers may choose from.  The options often pertain to the treatment of semi-structuredness (missing attributes, heterogeneous types, etc), where more than one sensible approaches are possible.

SQL++ is unifying: By appropriate choices of configuration options, the SQL++ semantics can morph into the semantics of existing semi-structured database query languages. The extensive experimental validation shows how SQL and four semi-structured database query languages (MongoDB, Cassandra CQL, Couchbase N1QL and AsterixDB AQL) are formally described by appropriate settings of the configuration options. 

Early adoption signs of SQL++ are positive: Version 4 of Couchbase's N1QL is explained as syntactic sugar over SQL++. AsterixDB will soon support the full SQL++ and Apache Drill is in the process of aligning with SQL++.
\end{abstract}

\section{Introduction}
\label{section:introduction}

Numerous databases marketed as SQL-on-Hadoop, NewSQL and NoSQL support Big Data applications. These databases generally support the 3Vs \cite{big-data-3v-oreilly-2012}. (i) Volume: amount of data (ii) Velocity: speed of data in and out (iii) Variety: semi-structured and heterogeneous data. Due to the Variety requirement, they have adopted semi-structured data models, which are generally different subsets of enriched JSON.%
\footnote{As explained below, the SQL data model itself is a subset of enriched JSON.}

Their evolving query languages fall short of full-fledged semi-structured query language capabilities%
\footnote{They also fall short of full-fledged SQL capabilities also.}
and have many variations. Some variations are due to superficial syntactic differences. However, other variations are genuine differences in query language capabilities and semantics. The lack of succinct, formal syntax and semantics inhibits a deep understanding of the various systems. It also impedes progress towards declarative languages for querying semi-structured data.

SQL++ is a semi-structured query language that is {\em backwards compatible} with SQL, in order to be easily understood and adopted by SQL programmers. The described semi-structured SQL++ data model is a superset of JSON and the SQL data model. The SQL++ model expands JSON with bags (as opposed to having JSON arrays only) and enriched values, i.e., atomic values that are not only numbers and strings (vendors have already adopted this extension \cite{bson}).
Vice versa, one may think of SQL++ as expanding SQL with JSON features: arrays, heterogeneity, and the possibility that any value may be an arbitrary composition of the array, bag and tuple constructors, hence enabling arbitrary nested structures, such as arrays of arrays. The SQL++ query language inputs and outputs SQL++ data. It makes the following contributions towards the evolution of query languages for JSON databases.

\noindent {\em Full-fledged semi-structured language} Many commercial JSON databases started as key-value and document-oriented databases. Others started with SQL as their base. In either case, they grow towards full-fledged JSON databases. SQL++ provides a full-fledged target language whose semantics pick the salient features of past full-fledged \textit{declarative} query languages for non-relational data models: OQL \cite{oql-dbpl-1989}, the nested relational model and query languages \cite{nest-unnest-pods-1982,nested-relational-vldb-1988,nested-relational-workshop-lncs-1989} and XQuery (and other XML-based query languages) \cite{xquery-3.0-w3c-2013,xml-ql-computer-networks-1999,xml-query-language-survey-sigmod-record-2000}. 
Importantly, in the spirit of XQuery and OQL, SQL++ is a fully composable and semi-structured language, hence being able to input and output nested and heterogeneous structures. The cornerstones of achieving this while also being SQL compatible, are described next.

\noindent {\em Removing restrictions of SQL semantics}  Instead of inventing (or re-inventing) many new features, the extension from SQL to SQL++ is most often achieved simply by {\em removing} semantic restrictions of SQL: 
\begin{compact_enum}
\item Unlike SQL's \texttt{FROM} clause variables, which bind to tuples only, the FROM clause variables of SQL++ bind to any JSON element.
\item SQL++ is fully composable in the sense that subqueries can appear anywhere, potentially creating nested results when they appear in the \texttt{SELECT} clause.
\item While correlation of the subqueries in a \texttt{FROM} clause is not allowed in SQL, the SQL++ subqueries of a \texttt{FROM} clause may be correlated with (earlier defined) variables of the same \texttt{FROM} clause.
\item The groups created by the \texttt{GROUP BY} are directly usable in nested queries - as opposed to SQL's approach where they may only participate in aggregate functions in very limited and particular ways. Interestingly, the SQL++ approach ends up explaining in a simpler way even plain SQL's grouping and aggregation.
\item Unlike SQL, the SQL++ semantics do not require schema or any homogeneity on the input data. 
\end{compact_enum}

\noindent {\em Extensions: Variables binding to attribute/value pairs and array positions} In addition to not requiring a schema, SQL++ can allow a pair of variables of the \texttt{FROM} clause to range over the attribute name and attribute value pairs of input tuples. Similarly, when a \texttt{FROM} clause ranges over an array, SQL++ allows a pair of variables to capture the data at an array position and the index of this position. In this way SQL++ seamlessly expands the logic of SQL to ``schema'' inspection and arrays.

As a result, the SQL++ syntax and semantics are short (as evidenced by this paper's syntax and semantics definitions), despite adding functionality on SQL. A key methodology that led to short, succinct semantics is the staging: First we define an {\em SQL++ core}. Consequently, SQL compatibility is achieved as syntactic sugar over the core. Similarly, many JSON database query language constructs of various vendors, can also be reduced to syntactic sugar over the core. As a proof of concept, we coauthored with the Couchbase N1QL team a reduction of the N1QL v4 query language into the SQL++ core, by using simple rewriting rules that treat the N1QL v4 syntactic constructs as syntactic sugar over SQL++. Indeed the formal semantics of N1QL v4 are explained by this reduction \cite{n1ql-to-sqlpp-2015}. (We had earlier collaborated with Couchbase to align N1QL v4 with SQL++.)  

The simplicity and power of the language have led to positive early adoption signs: (a) the Couchbase N1QL - SQL++ alignment, (b) an SQL++ interface for AsterixDB \cite{asterixdb-dpd-2011} captures the SELECT-FROM-WHERE language part and a full release of SQL++ for AsterixDB will soon follow, and (c) collaboration with MapR's Apache Drill towards aligning the query language with SQL++. An earlier version of SQL++ has been used by the federated query processor of the FORWARD application development platform \texttt{http://forward.ucsd.edu/sqlpp}.

\noindent {\em Configurable}  At the same time, via communication with the vendors we recognize that a model and query language standard is probably premature for such a young and fast-evolving area. Yet, the language designers need to know now the options that are available, especially as it pertains to the handling of semi-structured aspects (semantics for missing attributes, heterogeneous types, etc), which are not captured by the SQL backwards compatibility. 

To handle this requirement, SQL++ includes {\em configuration options} that describe 
\begin{compact_enum}
\item which features are supported and 
\item (for the supported features) different options of language semantics that formally capture the variations of existing database query languages. 
\end{compact_enum}

\noindent {\em Unifying and Explanatory} By appropriate choices of configuration options, the SQL++ semantics morph into the semantics of other SQL+JSON databases. The paper shows how the SQL standard and four well known (Cassandra CQL, MongoDB, Couchbase N1QL, AsterixDB AQL) semi-structured database query languages, are explained as particular settings of the configuration options. While SQL++ does not support the exact syntax of any of these four databases, it can be morphed by the configuration options to support equivalent queries. 
By understanding each database's capabilities in terms of SQL++ configuration options, the reader can focus on the fundamental differences of the databases without being confused by syntactic idiosyncracies of various query languages and superficial differences in the documented descriptions of their semantics. To further facilitate understanding of SQL++ and the effect of the various configuration options, we provide a web-accessible reference implementation of SQL++ at \texttt{http://forward.ucsd.edu/sqlpp}.

An earlier, extended version \cite{sqlpp-survey-2014} shows how an additional six databases correspond to particular settings of the configuration options. 
We expect that some of the results listed in the feature matrices describing configuration options will change in the next years as the space evolves rapidly.
Despite the forthcoming changes, we expect SQL++ to remain a standing contribution in
\begin{compact_enum}
\item guiding towards formal, SQL backwards compatible, minimalistic syntax and semantics for a full-fledged semi-structured JSON language and
\item enable understanding of differences in this space and guide towards rational decisions via the configuration option mechanism.
\end{compact_enum}

\section{Data Model}
\label{section:data-model}
The SQL++ data model is a superset of both SQL's relational tables and JSON, based on the observation that they both use similar concepts: A SQL tuple corresponds to a JSON object literal and a SQL string/integer/boolean to the respective JSON scalar. A JSON array is similar to a SQL table (bag) with order.

\begin{figure}[t]
\scriptsize
\begin{tabular}{|r|lrl|}
\hline
 1  & \gn{named\_value}         & \gp   & \gs{name} \gl{::} \gn{value} \\ \hline
 2  & \gn{value}                & \gp   & \gl{null} \\
 3  &                           & \gd   & \gl{missing} \\
 4  &                           & \gd   & \gn{scalar\_value} \\
 5  &                           & \gd   & \gn{complex\_value} \\
 6  & \gn{complex\_value}       & \gp   & \gn{tuple\_value} \\
 7  &                           & \gd   & \gn{collection\_value} \\ \hline
 8  & \gn{scalar\_value}        & \gp   & \gn{primitive\_value} \\
 9  &                           & \gd   & \gn{enriched\_value} \\
10  & \gn{primitive\_value}     & \gp   & \gl{'} \gs{string} \gl{'} \\
11  &                           & \gd   & \gs{number} \\
12  &                           & \gd   & \gl{true} \\
13  &                           & \gd   & \gl{false} \\
14  & \gn{enriched\_value}      & \gp   & \gs{type} \gl{(} (\gn{primitive\_value} \gl{,})+ \gl{)} \\ \hline
15  & \gn{tuple\_value}         & \gp   & \gl{\{} (\gs{name} \gl{:} \gn{value} \gl{,})+ \gl{\}} \\ \hline
16  & \gn{collection\_value}    & \gp   & \gn{array\_value} \\
17  &                           & \gd   & \gn{bag\_value} \\
18  & \gn{array\_value}         & \gp   & \gl{[} (\gn{value} \gl{,})* \gl{]} \\
19  & \gn{bag\_value}           & \gp   & \gl{\{\{} (\gn{value} \gl{,})* \gl{\}\}} \\
\hline
\end{tabular}
\caption{BNF Grammar for SQL++ Values}
\label{figure:values:bnf}
\end{figure}

\newcommand{\linevalues}[1]{%
    \IfEqCase*{#1}{%
    {named value}{(BNF line~1)}%
    {missing}{(BNF line~3)}%
    {scalar}{(BNF lines~8-9)}%
    {primitive}{(BNF lines~10-13)}%
    {enriched}{(BNF line~14)}%
    {tuple}{(BNF line~15)}%
    {collection}{(BNF lines~16-17)}%
    }[\errmessage{Unable to ref #1 for value BNF}]%
}%

\begin{figure}[t]
\includegraphics[width=\columnwidth]{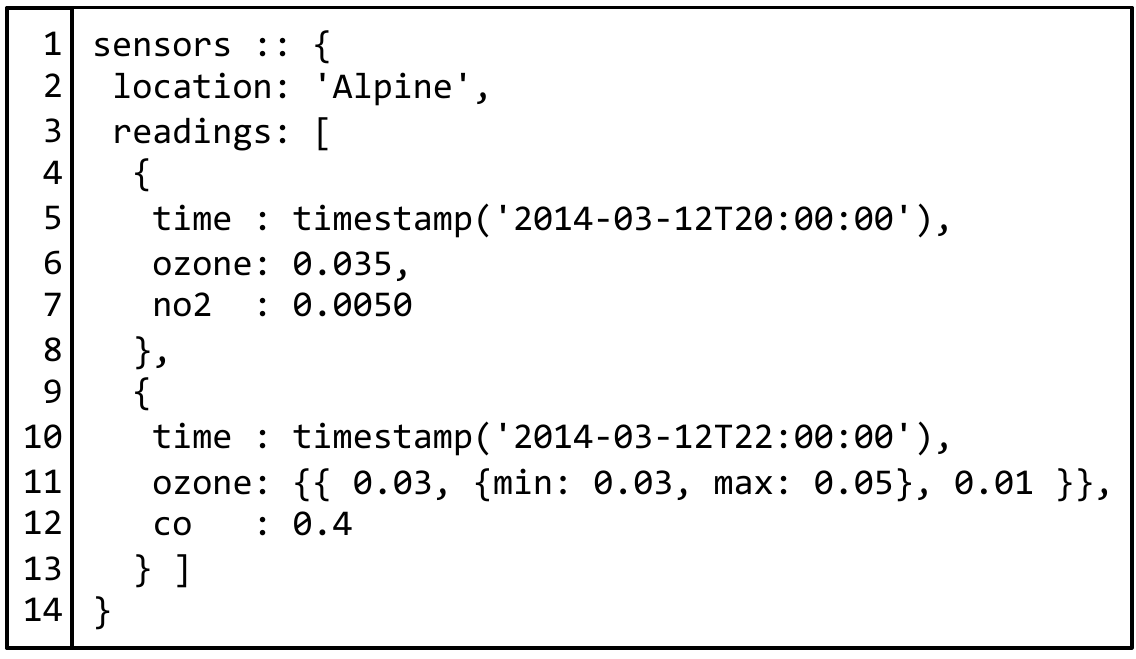}
\caption{Example of a SQL++ Value}
\label{figure:values:example-value}
\end{figure}

Figure~\ref{figure:values:bnf} shows the BNF grammar for SQL++ values, and Figure~\ref{figure:values:example-value} shows an example SQL++ value. A SQL++ database generally contains one or more SQL++ top-level \textit{named values} \linevalues{named value}. A \textit{name}, such as the name \texttt{sensors} of Figure~\ref{figure:values:example-value}, is a string and is unique. A value is a \textit{scalar}, \textit{complex}, \texttt{missing} (explained below) or \texttt{null}. A complex value is either a \textit{tuple} or a \textit{collection}. 
A tuple is a set of attribute name/value pairs, where each name is a unique string within the tuple (as in SQL). Notice, SQL++ syntax follows JSON. Therefore, tuples are denoted by \texttt{\{} \texttt{\}}. In Figure~\ref{figure:values:example-value}, \texttt{sensors} is a tuple with attributes \texttt{location} and \texttt{readings}.  Each attribute value can be a scalar, a \texttt{null}, or a complex value \linevalues{tuple}.

A collection is either an \textit{array} or a \textit{bag} \linevalues{collection}. Both arrays and bags may contain duplicate elements. An array is ordered (similar to a JSON array) and each element is accessible by its ordinal position. (See specifics of access by position in Section~\ref{section:paths}.) In contrast, a bag is unordered (similar to a SQL table) and its elements cannot be accessed by ordinal position. Following JSON's syntax, arrays are denoted with \texttt{[...]}, whereas following AQL's syntax, bags are denoted with \texttt{\{\!\{...\}\!\}}. The value of \texttt{readings} is an array (Example lines~3-13). The value of \texttt{ozone} is a bag (Example line~11). 

A scalar value is either \textit{primitive} or \textit{enriched} \linevalues{scalar}. Primitive values are the scalar values of the JSON specification, i.e. strings, numbers or booleans \linevalues{primitive}. Enriched values (such as dates and timestamps) are extensions over JSON, and are specified using a type constructor over primitives \linevalues{enriched}. The value \texttt{timestamp(`2014-03-12T20:00:00')} (Example line 5) is enriched. 

The elements of an array/bag can be any kind of value and can be heterogeneous. That is, there are no restrictions between the elements of an array/bag. For example, the two tuples in the array of lines~3 to~13 are \textit{heterogeneous} because: (i) each tuple has a different set of attributes (lines~5-7 vs 10-12), (ii) the \texttt{ozone} attribute of the first tuple maps to a number (line~6) while the \texttt{ozone} of the second maps to a bag (line~ 11). The elements of the \texttt{ozone} bag of line~11 are also heterogeneous (1st and 3rd are numbers, 2nd is a tuple). Furthermore, unlike SQL where the values are tables that have homogeneous tuples that have scalars, SQL++ allows {\em arbitrary composition of complex values}. E.g., the top level value is a tuple, the array of lines~3-13 has two heterogeneous tuples, and the bag of line~11 has two scalars and one tuple.

SQL++ and many semi-structured databases also contains maps. A map contains mappings, where each mapping maps a left value to a right value, and each left value is unique within the map. In effect, both a map and a tuple are sets of name/value pairs where names are unique. Their only difference is that a map allows the left value of each mapping to be any arbitrary value, whereas a tuple restricts attribute names to strings (as in SQL). Therefore we do not further discuss maps, as they are essentially tuples without the string restriction.

\section{Queries, Environments and Binding tuples}
\label{section:environment-and-sfw}

\begin{figure}[t!]
\scriptsize
\begin{tabular}{|@{~}rc@{~}l@{~}|}
\hline
\hline
         1  & \multicolumn{2}{@{}l@{~}|}{\gn{query}} \\
         2  & \gp & \gn{sfw\_query} \\
         3  & \gd & \gn{expr\_query} \\
         4  & \gd & \gn{config\_annotation}+ \gl{(} \gn{query} \gl{)} \\ \hline
         5  & \multicolumn{2}{@{}l@{~}|}{\gn{sfw\_query}} \\
         6  & \gp & \gn{select\_clause} (see Figure~\ref{figure:select:bnf}) \\
         7  &     & \gn{from\_clause} (see Figure~\ref{figure:from:bnf}) \\
         8  &     & (\gl{WHERE} \gn{expr\_query})? \\
         9  &     & (\gl{GROUP BY} \gn{expr\_query} (\gl{AS} \gn{var})? \\ 
        10  &     & ~~~~(\gl{,} \gn{expr\_query} (\gl{AS} \gn{var})?)*)? \\
        11  &     & (\gl{HAVING} \gn{expr\_query})? \\
        12  &     & ((\gl{UNION}\gd\gl{INTERSECT}\gd\gl{EXCEPT}) \gl{ALL}? \gn{sfw\_query})? \\
        13  &     & (\gl{ORDER BY} \gn{expr\_query} (\gl{ASC}\gd\gl{DESC})? \\
        14  &     & ~~~~(\gl{,} \gn{expr\_query} (\gl{ASC}\gd\gl{DESC})?)*)? \\
        15  &     & (\gl{LIMIT} \gn{expr\_query})? \\
        16  &     & (\gl{OFFSET} \gn{expr\_query})? \\ \hline
        17  & \multicolumn{2}{@{}l@{~}|}{\gn{expr\_query}} \\
        18  & \gp & \gl{(} \gn{sfw\_query} \gl{)} \\
        19  & \gd & \gn{named\_value} \\
        20  & \gd & \gn{var} \\
        21  & \gd & \gn{expr\_query} \gl{.} \gs{attr\_name} \\
        22  & \gd & \gn{expr\_query} \gl{[} \gs{expr\_query} \gl{]} \\
        23  & \gd & \gn{function\_name} \gl{(}\gn{expr\_query}? (\gl{,} \gn{expr\_query})* \gl{)} \\
        24  & \gd & \gl{\{} (\gs{attr\_name}\gl{:}\gn{expr\_query})? (\gl{,} \gs{attr\_name}\gl{:}\gn{expr\_query})* \gl{\}} \\
        25  & \gd & \gl{[} \gn{expr\_query}? (\gl{,} \gn{expr\_query})* \gl{]} \\
        26  & \gd & \gl{\{\{} \gn{expr\_query}? (\gl{,} \gn{expr\_query})* \gl{\}\}} \\
        27  & \gd & \gs{value} \\ \hline
        28  & \multicolumn{2}{@{}l@{~}|}{\gn{config\_annotation}} \\
        29  & \gp & \gl{@tuple\_nav} \gl{\{}\gn{path\_param} (\gl{,} \gn{path\_param})*\gl{\}} (see Figure~\ref{figure:paths:bnf}) \\
        30  & \gd & \gl{@array\_nav} \gl{\{}\gn{path\_param} (\gl{,} \gn{path\_param})*\gl{\}} (see Figure~\ref{figure:paths:bnf}) \\
        31  & \gd & \gl{@eq} \gl{\{}\gn{eq\_param} (\gl{,} \gn{eq\_param})*\gl{\}} (see Figure~\ref{figure:equality:bnf}) \\
        32  & \gd & \gl{@from} \gl{\{}\gn{from\_param} (\gl{,} \gn{from\_param})*\gl{\}} (see Figure~\ref{figure:from:bnf}) \\
        33  & \gd & \gl{@sql} \\
        34  & \gd & \ldots \\
\hline
\hline
\end{tabular} 
\caption{BNF Grammar for SQL++ Queries}
\label{figure:query:bnf}
\end{figure}

\newcommand{\linequery}[1]{%
    \IfEqCase*{#1}{%
    {sfw query}{lines~5-16}%
    {expression query}{lines~17-27}%
    {configuration options}{(line~4)}%
    {configured query}{(Figure~\ref{figure:query:bnf}, lines~4, 28-34)}%
    {parameter group}{(lines~28-32)}%
    {config annotation}{(line~4)}%
    {macro}{(line~33)}%
    {tuple nav}{(Figure~\ref{figure:query:bnf}, line~21)}%
    {array nav}{(line~22)}%
    {path parameter groups}{(Figure~\ref{figure:query:bnf}, lines~29-30)}%
    {equality parameter group}{(Figure~\ref{figure:query:bnf}, line~31)}%
    {expression, sql}{(Figure~\ref{figure:query:bnf}, lines~18-19)}%
    {expression, extensions}{(Figure~\ref{figure:query:bnf}, lines~20-27)}%
    {constructors}{(Figure~\ref{figure:query:bnf}, lines~24-26)}%
    {group by}{(Figure~\ref{figure:query:bnf}, lines~9-10)}%
    }[\errmessage{Unable to ref #1 for query BNF}]%
}%

\begin{figure}[t!]
\includegraphics[width=\columnwidth]{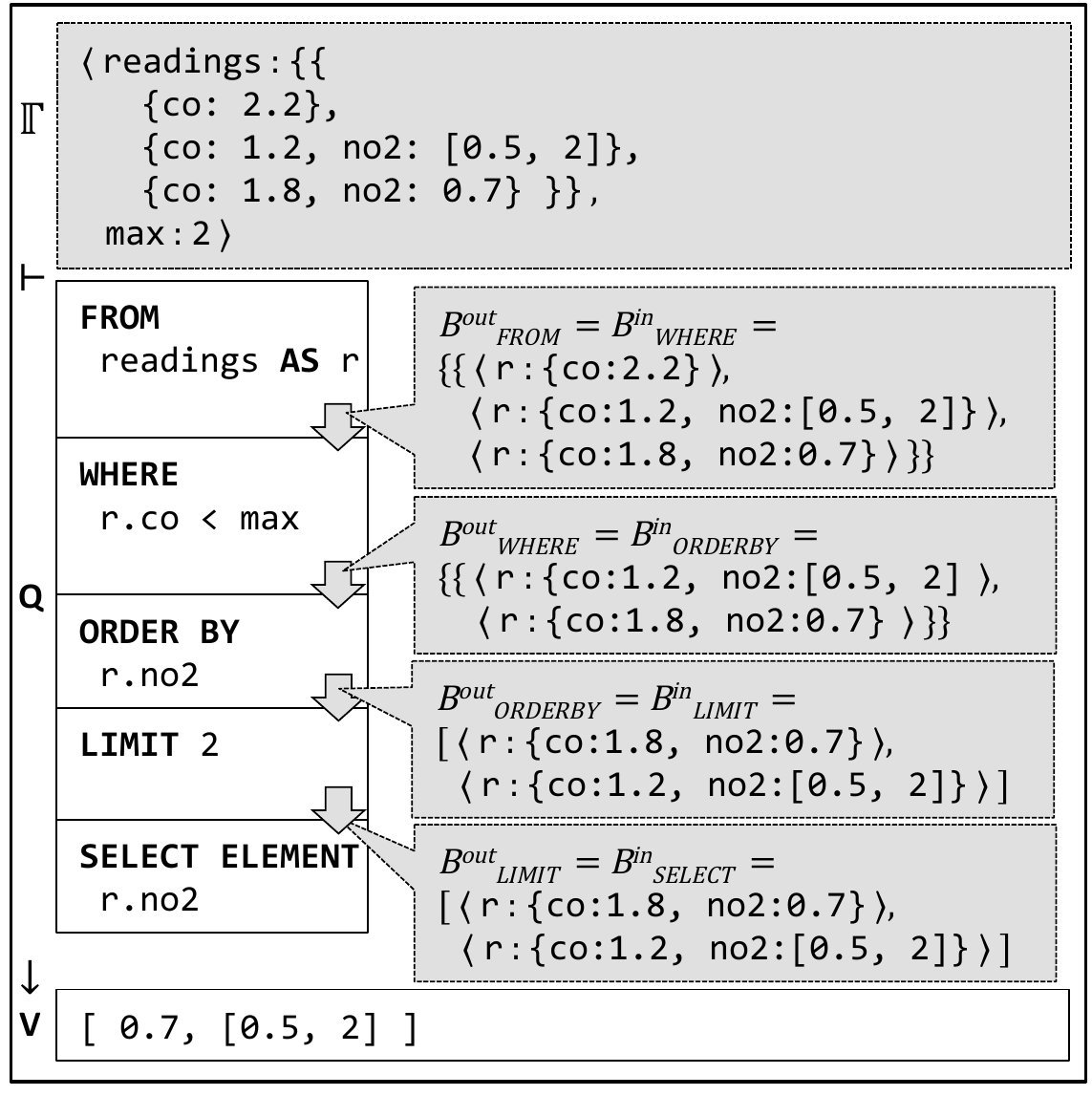}
\caption{Binding tuples in an SFW Query}
\label{figure:sfw:example-sfw}
\end{figure} 

A SQL++ query is either an \textit{SFW query} (i.e. \texttt{SELECT-FROM-WHERE}, \linequery{sfw query} of the grammar of Figure \ref{figure:query:bnf}) or an \textit{expression query} (or \textit{expression},  \linequery{expression query}). Unlike SQL expressions, which are restricted to outputting scalar and null values, SQL++ expression queries output arbitrary values, and are fully composable with SFW queries. Furthermore, SQL++ supports the top-level query to also be an expression query, not just a SFW query as in SQL. 

A SQL++ (sub)query is evaluated within an environment, which provides variable bindings (as defined next) and optionally configuration options, defined in Section~\ref{section:configs}.

\highlight{Environments and Variable Bindings} Each SQL++ (sub)query $q$ is evaluated within an \textit{environment} (for brevity, \textit{env}) $\env$ comprising a \textit{config env} $\env_c$ and a \textit{variable binding env} $\env_b$ (or \textit{binding env}). (We may omit the subscript $b$ in examples where the environment is clearly a variable binding environment.) A binding env is a \textit{binding tuple} $\langle x_1 : v_1, \ldots, x_n : v_n \rangle$, where each $x_i$ is a variable name that is unique, and binds to the SQL++ value $v_i$. Notice that a binding tuple is structurally a SQL++ tuple. The characterization ``binding'' pertains to its use in the semantics, rather than structural differences.

Given two binding tuples $b$ and $b'$, their concatenation is a binding tuple, denoted as $b || b'$, that has the variable bindings of both $b$ and $b'$. In order to define variable scoping later, we require that if a variable $x$ is bound in both $b$ and $b'$, $b || b'$ retains only the binding of $x$ in $b$. Hence each variable remains unique within a binding tuple.

The notation $\env_b \vdash q \rightarrow v$ denotes that the SQL++ query $q$ evaluates to the value $v$ within $\env_b$, i.e. when every free variable of $q$ is instantiated by its binding in $\env_b$. For example, $\langle x: 5, y:3 \rangle \vdash (x + y)/2 \rightarrow 4$.

The binding environments of a query and of its subqueries are produced as follows:
\begin{compact_enum} 
\item When a query is submitted to a database, it is evaluated in the environment comprising $\env_b = \langle n_1:v_1,\ldots,n_m:v_m \rangle$ where the pairs $n_i:v_i$ are the database's named values $v_i$ and their names $n_i$. For example, Figure~\ref{figure:sfw:example-sfw} shows a SFW query that filters for \texttt{co} readings less than \texttt{max}, and outputs the top-2 \texttt{no2} readings. The query is evaluated on a database that has two named values: the bag named \texttt{readings} and the number named \texttt{max}. Thus, the query's environment $\env_b$ has variable bindings \texttt{readings} and \texttt{max}.
\item The \texttt{FROM} clause of a SFW query produces new environments by concatenating bindings of its variables to the environment of its SFW query, as explained below. Subqueries are evaluated in these new environments. (The \texttt{GROUP BY} clause   also produce additional variable bindings, as explained in Section~\ref{section:group-by} .)
\end{compact_enum}

\highlight{SFW Query Clauses} Similar to SQL semantics, the clauses of a SQL++ SFW query are evaluated in the following order: \texttt{FROM}, \texttt{WHERE}, \texttt{GROUP BY}, \texttt{HAVING}, \texttt{ORDER BY}, \texttt{LIMIT / OFFSET} and \texttt{SELECT}%
\footnote{SQL++ also supports a syntax improvement where \texttt{SELECT} is optionally written as the last clause.}.
The \texttt{FROM} expression (\texttt{readings} in Figure~\ref{figure:sfw:example-sfw}) is evaluated in the environment of the query ($\env$ in the example). Thereafter the \texttt{FROM} clause outputs the bag of binding tuples $B^{out}_{\texttt{FROM}}$ (which has 3 binding tuples in the example). In each binding tuple of $B^{out}_{\texttt{FROM}}$, each variable of the \texttt{FROM} clause is bound to a value. There are no restrictions that a variable binds to homogenous values across binding tuples. In the example, \texttt{r} respectively binds to 3 values that are heterogeneous.

Each subsequent clause inputs a bag of binding tuples, evaluates the constituent expressions of the clause (which may themselves contain nested SFW queries), and outputs a bag of binding tuples that is in turn input by the next clause. For instance, the \texttt{WHERE} clause inputs the bag of binding tuples that have been output by \texttt{FROM} ($B^{out}_{\texttt{FROM}}$ = $B^{in}_{\texttt{WHERE}}$), and outputs the subset thereof that satisfies the condition expression $e$ of the \texttt{WHERE} clause. Condition $e$ is evaluated once for each input binding tuple $b$ in the binding env $b \| \env_b$, where $\env_b$ is the binding env of the SFW query. In the example, the condition \texttt{r.co < max} is evaluated once for each of the 3 input binding tuples of $B^{in}_{\texttt{WHERE}}$. The binding env of the first evaluation is $\langle \texttt{r} : \texttt{\{co:2.2\}} \rangle || \env_b = \langle \texttt{r} : \texttt{\{co:2.2\}}, \texttt{readings}:\ldots, \texttt{max}:\texttt{2} \rangle$, thus the first binding tuple is not output.

The pattern of ``input bag, evaluate constituent expressions, output bag'' has a few exceptions, each of them exemplified in the example: First, the \texttt{ORDER BY} clause inputs a bag of binding tuples and outputs an array of binding tuples. Second, a \texttt{LIMIT}/\texttt{OFFSET} clause need not evaluate its constituent expression for each input binding tuple. Finally, the \texttt{SELECT ELEMENT} clause inputs a bag (resp. array, if \texttt{ORDER BY} is present) of binding tuples, and outputs the SFW query's result, which is a bag (resp. array) with exactly one element for each input binding tuple. In the example, the \texttt{SELECT ELEMENT} expression \texttt{r.no2} is evaluated once for each of the 2 input binding tuples of 
$B^{in}_{\texttt{SELECT}}$. The first evaluation is within the binding env $\langle \texttt{r} : \texttt{\{co:1.8,no2:0.7\}} \rangle || \env_b$, and outputs the element \texttt{0.7}.

Finally, notice that the above discussion of SFW queries did not capture the set operators \texttt{UNION}, \texttt{INTERSECT} and \texttt{EXCEPT}. As is the case with SQL semantics too, the coordination of \texttt{ORDER BY} with the set operators is treated as a special case, which is described in the extended version.

\highlight{Generalizing Variable Bindings} Assume that \texttt{readings} in Figure~\ref{figure:sfw:example-sfw} were a SQL table (i.e. a bag of homogeneous tuples) with number attributes \texttt{co} and \texttt{no2}. SQL semantics dictate that \texttt{FROM readings AS r} outputs homogenous tuples with attributes \texttt{co} and \texttt{no2}. In contrast, the SQL++ \texttt{FROM} clause outputs binding tuples, each of which binds the single variable \texttt{r} to a potentially heterogeneous tuple. In the example, \texttt{r} binds to tuples containing \texttt{co} and \texttt{no2} attributes. This additional level of nesting generalizes SQL++ variable bindings beyond SQL's homogeneous tuples, to encompass arbitrary \texttt{readings} elements that are heterogeneous (as shown in the example), including a mixture of scalars, tuples and/or collections.  

\highlight{Composability} Notice that query evaluation treats each value named $n$ identically to a variable named $n$. For example, the query of Figure~\ref{figure:sfw:example-sfw} could be a nested query where \texttt{readings} and \texttt{max} are variables defined in the \texttt{FROM} clause of the parent query. This property serves the composability of the language.

\highlight{Heterogeneity} SQL++ supports named values, query results and variables to be arbitrary values. This is in contrast to SQL which restricts named values to flat homogeneous tables (created by \texttt{CREATE TABLE}, \texttt{CREATE VIEW} and \texttt{WITH}), query results to flat homogeneous tables, and tuple variable bindings to homogeneous flat tuples. 

The combination of composability and heterogeneity are needed to input and output arbitrarily structured SQL++ values.

\section{Configuration Options}
\label{section:configs}

Config options enable SQL++ to be instantiated to different capabilities and semantics. In effect, they represent the choices that language designers may consider. Some options specify whether a syntactic feature is supported or not. For example, \texttt{GROUP BY} is supported by MongoDB, but not Cassandra. Other options specify variations in the semantics of a syntactic feature. For example, given a tuple \texttt{t} with a single attribute \texttt{a}, AsterixDB evaluates the path step \texttt{t.b} to return \texttt{null} (effectively losing the difference between a missing \texttt{b} attribute and a \texttt{b} attribute with \texttt{null} value), Couchbase returns \texttt{missing} (which behaves differently from \texttt{null}), and SQL throws an \texttt{error}.   

Config options are extensively utilized throughout SQL++. This section presents the formalism, a simple use case in path navigation (Section~\ref{section:paths}) and a complex use case in equality (Section~\ref{section:equality}). The support/non-support of query features described in Section~\ref{section:clauses} are also specified by config options. Section~\ref{section:experiments} experimentally validates these config options by using them to systematically (a) describe the features of the surveyed databases as a subset of SQL++ and (b) describe their semantics as a specific instantiation of SQL++ semantics.

\subsection{Config Envs, Parameters and Options}
\label{section:parameters}
 
A \textit{config annotation} $\texttt{@}g \texttt{\{} p \texttt{:} o \texttt{\}(} q \texttt{)}$ \linequery{configured query} configures (sub)query $q$ with \textit{parameter group} $g$, \textit{parameter name} $p$ and option $o$.%
\footnote{As syntactic sugar, parameters of the same group only need to specify the group once \linequery{parameter group}, i.e. $\texttt{@}g \texttt{\{} n_1 \texttt{:} o_1  \texttt{,} \ldots \texttt{,} n_m \texttt{:} o_m \texttt{\}(} q \texttt{)}$ is sugar for \\
$\texttt{@}g \texttt{\{} n_1 \texttt{:} o_1 \texttt{\}(} \ldots \texttt{(@}g \texttt{\{} n_m \texttt{:} o_m \texttt{\}(} q \texttt{)} \ldots \texttt{)}$. Consecutive config annotations can also omit parentheses \linequery{config annotation}.
} 
The config annotation is statically specified: it cannot be the result of another query. 
A query is evaluated in the configuration env(ironment) created by its config annotation. A config env $\env_c$ is a config tuple $\langle p_1 : o_1, \ldots, p_m : o_m \rangle$, where each parameter $p_i$ maps to an option $o_i$.  

Using config options, language designers have the flexibility for very fine-grained customization of language semantics. To assist designers in choosing sets of options that are compatible with each other a \textit{macro config annotation} $\texttt{@}g \texttt{\{} p \texttt{:} o \texttt{\}}$ expands to a set of coordinated config annotations $\texttt{@}g \texttt{\{} p_1 \texttt{:} o_1 \texttt{\}} \ldots \texttt{@}g \texttt{\{} p_m \texttt{:} o_m \texttt{\}}$. For example, a macro config annotation may dictate that the semantics of \texttt{WHERE} in the presence of \texttt{null} and \texttt{missing} are identical to SQL's 3-valued logic, whereas \texttt{missing} is treated as \texttt{null}. Alternately, the annotation may dictate another form of semantics where the distinction between \texttt{null} (i.e., unknown) and \texttt{missing} (i.e. inapplicable) is retained.

\highlight{The utility of configuration override in federated query processing} A query's configuration environment may be overriden by annotations that apply to its subqueries. Suppose $\texttt{@}g \texttt{\{} p \texttt{:} o \texttt{\}(} q \texttt{)}$ is evaluated in env $\env = (\env_b, \env_c)$. (The query $q$ may be a subquery of an enclosing query.) Then $q$ is evaluated in env $\env' = (\env_b, \langle \texttt{@}g\texttt{.}p : o \rangle \| \env_c)$. That is, $q$ is evaluated in a context where $p$ is set to $o$, regardless of whether it could have been configured differently in an enclosing query. Similar to binding tuples, given two config tuples $c$ and $c'$, their concatenation $c \| c'$ retains the mapping of $c$ if parameter $p$ is mapped in both $c$ and $c'$. Therefore, two different parts of a query may correspond to configurations of different databases. 

The ability to configure subqueries with different annotations becomes handy when SQL++ is used as a federated query language. In such case, if a subquery that is pushed to a certain database is configured according to the options of such database then it can indeed be pushed. In effect, the task of a federated query processor (such as FORWARD) is to rewrite the user query into one where all subqueries that are pushed to an underlying database hav been configured according to the configuration options of such database.

\subsection{Path Navigation}
\label{section:paths}

\begin{figure}
\scriptsize
\begin{tabular}{|@{~}rc@{~}l@{~}|}
\hline
\hline
         1  & \multicolumn{2}{@{}l@{~}|}{\gn{path\_param}} \\
         2  & \gp   & \gl{absent\phantom{\_}~~~~~~:}    (\gl{null}\gd\gl{missing}\gd\gl{error}) \\
         3  & \gd   & \gl{type\_mismatch:}              (\gl{null}\gd\gl{missing}\gd\gl{error}) \\
\hline
\hline
\end{tabular} 
\caption{BNF Grammar for Path Parameters/Options}
\label{figure:paths:bnf}
\end{figure}

\newcommand{\linepath}[1]{%
    \IfEqCase*{#1}{%
    {absent}{(Figure~\ref{figure:paths:bnf}, line~2)}%
    {type mismatch}{(line~3)}%
    }[\errmessage{Unable to ref #1 for query BNF}]%
}%

\begin{figure}
\begin{mdframed}[innerleftmargin=0.2cm]
\scriptsize
\[
\begin{array}{@{}r@{~}l}
    t\texttt{.}a            &
        \rightarrow
        \begin{cases}
        v                                   & \text{if } t \text{ is a tuple that}  \\
                                            & \text{maps } a \text{ to } v          \\
        \texttt{@tuple\_nav.absent}         & \text{if } t \text{ is a tuple that}  \\
                                            & \text{does not map } a                \\
        \texttt{@tuple\_nav.type\_mismatch} & \text{otherwise}
        \end{cases}         \\ & \\
    a\texttt{[}i\texttt{]}  &
        \rightarrow
        \begin{cases}
        v                                   & \text{if } a \text{ is an array with}         \\
                                            & i\text{-th element as } v                     \\
        \texttt{@array\_nav.absent}         & \text{if } a \text{ is an array with } n      \\
                                            & \text{elements} ~ \wedge ~ (i < 1 \vee i > n) \\
        \texttt{@array\_nav.type\_mismatch} & \text{otherwise}
        \end{cases}         \\ & \\
\end{array}
\]
\normalsize
\end{mdframed}
\caption{Path Evaluation Functions for $t.a$ and $a[i]$}
\label{figure:paths:semantics}
\end{figure}

\begin{figure}
\includegraphics[width=\columnwidth]{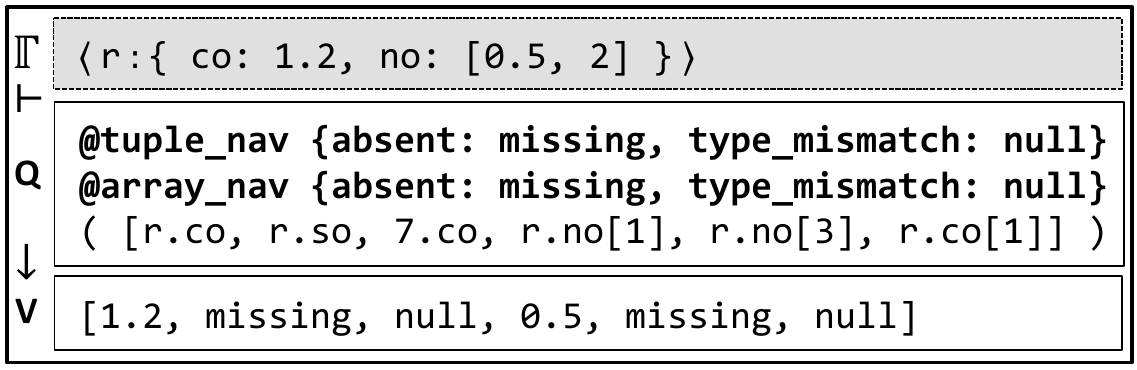}
\caption{Configuring Tuple/Array Navigation}
\label{figure:paths:example-paths}
\end{figure}

A \textit{tuple path navigation} \texttt{t.a} from the tuple \texttt{t} to its attribute \texttt{a} \linequery{tuple nav} returns the value of \texttt{a}. An \textit{array path navigation} \texttt{a[i]} returns the \texttt{i}-th element of the array \texttt{a} \linequery{array nav}. Notice that consecutive tuple/array navigations (e.g. \texttt{r.no[1]}) navigate deeply into complex values, and each navigation evaluates to a unique value. This is identical to SQL's behavior, but different from XQuery/XPath's behavior where each path step may return a set of one or more nodes. 

Since SQL++ does not assume a schema, the semantics must also specify the return value when \texttt{t} is not a tuple or when it is a tuple that does not have an \texttt{a} attribute. Multiple options are possible for these cases and, indeed, multiple options have been taken by the surveyed databases. Figure~\ref{figure:paths:semantics} specifies the evaluation function of $t.a$ and the function of $a[i]$, which utilize parameters from the \texttt{@tuple\_nav} and \texttt{array\_nav} parameter groups \linequery{path parameter groups}. The \texttt{absent} parameter \linepath{absent} specifies the returned value when an attribute/array element is absent: \texttt{null}, \texttt{missing}, or throw an error. The \texttt{type\_mismatch} parameter \linepath{type mismatch} specifies whether to return \texttt{null}, \texttt{missing}, or throw an error when a tuple/array navigation is invoked on a non-tuple/array. 
For brevity, we omit trivial conversions from a parameter option to a SQL++ value. For example, $t.a \rightarrow \texttt{@tuple\_nav.absent}$ denotes returning a SQL++ \texttt{null} value for the parameter option \texttt{null}, a \texttt{missing} value for the option \texttt{missing}, and throwing an error for the option \texttt{error}.

In the query of Figure~\ref{figure:paths:example-paths} the \texttt{@tuple\_nav}/\texttt{@array\_nav} config annotations specify that navigating into an absent attribute/element returns \texttt{missing}, and navigating from a non-tuple/array returns \texttt{null}. As another example, SQL's semantics for tuple navigation is specified by \texttt{@tuple\_nav \{absent: error, type\_mismatch: error\}}. Since a SQL database has fixed schemas, these errors are caught by static type checking.

\begin{table}[t!]
\centering
\scriptsize
\begin{tabular}{|l|l||l||l|}
\hline
    \gl{@nav.failure\phantom{\_\_}~~~~~~~~~~:}  & \gl{error}        & \gl{null}         & \gl{missing}      \\ \hline \hline
\texttt{@tuple\_nav.absent\phantom{\_}~~~~~~:}  & \texttt{error}    & \texttt{null}     & \texttt{missing}  \\
\texttt{@tuple\_nav.type\_mismatch:}            & \texttt{error}    & \texttt{null}     & \texttt{missing}  \\
\texttt{@array\_nav.absent\phantom{\_}~~~~~~:}  & \texttt{error}    & \texttt{null}     & \texttt{missing}  \\
\texttt{@array\_nav.type\_mismatch:}            & \texttt{error}    & \texttt{null}     & \texttt{missing}  \\
\hline
\end{tabular}
\caption{\gl{@nav.failure} Macro Expansion}
\label{table:paths:macro}
\end{table}

Although the four path parameters are independent of each other and can (in principle) be configured with different options, we recommend language designers to utilize the macro config annotation \gl{@nav.failure} so that path navigation behaves consistently for all navigation failures. The macro supports three options, which expand to parameters and respective options as specified in Table~\ref{table:paths:macro}: (i) \gl{error} specifies a strict form of path navigation (as in SQL), such that navigation failures result in errors thrown. Notice, it expands to \texttt{@tuple\_nav.absent:error}, \texttt{@tuple\_nav.type\_mismatch:error}, etc. (ii) \gl{null} specifies a lenient form of path navigation, such that failures result in \texttt{null} values (iii) likewise, \gl{missing} specifies lenient path navigation where failures result in \texttt{missing} values.

\subsection{Equality}
\label{section:equality}

\begin{figure}[t!]
\scriptsize
\begin{tabular}{|@{~}rc@{~}l@{~}|}
\hline
\hline
         1  & \multicolumn{2}{@{}l@{~}|}{\gn{eq\_param}} \\
         2  & \gp   & \gl{complex\phantom{\_\_}~~~~~~~~~~:} (\gl{yes~~~~~~}\phantom{\gd\gd\gl{null}\gd\gl{missing}}\gd\gl{error}) \\
         3  & \gd   & \gl{type\_mismatch\phantom{\_}~~~~~:} (\phantom{\gd}\gl{~~~~false}\gd\gl{null}\gd\gl{missing}\gd\gl{error}) \\ \hline
         4  & \gd   & \gl{null\_eq\_null~~~~~~~:}           (\gl{true}\gd\gl{false}\gd\gl{null}\gd\gl{missing}\gd\gl{error}) \\
         5  & \gd   & \gl{null\_eq\_missing~~~~:}           (\gl{true}\gd\gl{false}\gd\gl{null}\gd\gl{missing}\gd\gl{error}) \\
         6  & \gd   & \gl{null\_eq\_value ~~~~~:}           (\gl{true}\gd\gl{false}\gd\gl{null}\gd\gl{missing}\gd\gl{error}) \\
         7  & \gd   & \gl{missing\_eq\_missing~:}           (\gl{true}\gd\gl{false}\gd\gl{null}\gd\gl{missing}\gd\gl{error}) \\
         8  & \gd   & \gl{missing\_eq\_value~~~:}           (\gl{true}\gd\gl{false}\gd\gl{null}\gd\gl{missing}\gd\gl{error}) \\ \hline
         9  & \gd   & \gl{null\_and\_true~~~~~~:}           (\gl{true}\gd\gl{false}\gd\gl{null}\gd\gl{missing}\gd\gl{error}) \\
        10  & \gd   & \gl{null\_and\_null~~~~~~:}           (\gl{true}\gd\gl{false}\gd\gl{null}\gd\gl{missing}\gd\gl{error}) \\
        11  & \gd   & \gl{null\_and\_missing~~~:}           (\gl{true}\gd\gl{false}\gd\gl{null}\gd\gl{missing}\gd\gl{error}) \\
        12  & \gd   & \gl{missing\_and\_true~~~:}           (\gl{true}\gd\gl{false}\gd\gl{null}\gd\gl{missing}\gd\gl{error}) \\
        13  & \gd   & \gl{missing\_and\_missing:}           (\gl{true}\gd\gl{false}\gd\gl{null}\gd\gl{missing}\gd\gl{error}) \\
\hline
\hline
\end{tabular} 
\caption{BNF Grammar for Equality Parameters/Options}
\label{figure:equality:bnf}
\end{figure}

\def\andoperator{\mathbin{\circ_{\texttt{and}}}}

\begin{figure}[t!]
\begin{mdframed}[innerleftmargin=0cm]
\newlength{\eqcaselength}
\setlength{\eqcaselength}{2.85cm}
\scriptsize
\[
\begin{array}{@{}r@{~}l}
x \texttt{=} y              &
    \rightarrow
    \begin{cases}
        \makebox[\eqcaselength][l]{%
        $f_{\texttt{complex}}(x, y)$}       & \text{if } x \text{ or } y \text{ is complex}                 \\
                                            & \wedge ~ \texttt{@eq.complex} \text{ is } \texttt{yes}        \\
        \texttt{error}                      & \text{if } x \text{ or } y \text{ is complex}                 \\
                                            & \wedge ~ \texttt{@eq.complex} \text{ is } \texttt{error}      \\
        \texttt{@eq.null\_eq\_null}         & \text{if } x \text{ and } y \text{ are } \texttt{null}        \\
        \texttt{@eq.null\_eq\_missing}      & \text{if } x \text{ (resp. } y \text{) is } \texttt{null}     \\
                                            & \wedge ~ y \text{ (resp. } x \text{) is } \texttt{missing}    \\
        \texttt{@eq.null\_eq\_value}        & \text{if } x \text{ (resp. } y \text{) is } \texttt{null}     \\
                                            & \wedge ~ y \text{ (resp. } x \text{) is scalar}               \\
        \texttt{@eq.missing\_eq\_missing}   & \text{if } x \text{ and } y \text{ are } \texttt{missing}     \\
        \texttt{@eq.missing\_eq\_value}     & \text{if } x \text{ (resp. } y \text{) is } \texttt{missing}  \\
                                            & \wedge ~ y \text{ (resp. } x \text{) is scalar}               \\
        f_{\texttt{scalar}}(x, y)           & \text{if } x \text{ and } y \text{ are scalar}                \\
    \end{cases}             \\ & \\
f_{\texttt{scalar}}(x, y)   &
    \rightarrow
    \begin{cases}
        \makebox[\eqcaselength][l]{%
        $f_{\texttt{sql}}(x, y)$}           & \text{if } x \text{ and } y \text{ are strings}               \\
                                            & \text{(resp. nums, booleans)}                                 \\
        \texttt{@eq.type\_mismatch}         & \text{otherwise}                                              \\
    \end{cases}             \\ & \\ \hline
f_{\texttt{complex}}(x, y)  &
    \rightarrow
    \begin{cases}
        \makebox[\eqcaselength][l]{%
        $f_{\texttt{array}}(x, y)$}         & \text{if } x \text{ and } y \text{ are arrays}                \\
        f_{\texttt{bag}}(x, y)              & \text{if } x \text{ and } y \text{ are bags}                  \\
        f_{\texttt{tuple}}(x, y)            & \text{if } x \text{ and } y \text{ are tuples}                \\
        \texttt{@eq.null\_eq\_value}        & \text{if } x \text{ (resp. } y \text{) is } \texttt{null}     \\
                                            & \wedge ~ y \text{ (resp. } x \text{) is complex}              \\
        \texttt{@eq.missing\_eq\_value}     & \text{if } x \text{ (resp. } y \text{) is } \texttt{missing}  \\
                                            & \wedge ~ y \text{ (resp. } x \text{) is complex}              \\
        \texttt{@eq.type\_mismatch}         & \text{otherwise}                                              \\
    \end{cases}             \\ & \\
f_{\texttt{array}}(x, y)    &
    \rightarrow
    \begin{cases}
        \makebox[\eqcaselength][l]{%
        $x\texttt{[}1\texttt{]} \texttt{=} y\texttt{[}1\texttt{]} ~ \andoperator$}          & \text{if } x \text{ and } y \text{ each has }         \\
        ~~ \ldots \andoperator ~ x\texttt{[}n\texttt{]} \texttt{=} y\texttt{[}n\texttt{]}   & ~~ n \text{ elements}                                 \\
        \texttt{false}                                                                      & \text{otherwise}                                      \\
    \end{cases}             \\ & \\
f_{\texttt{bag}}(x, y)      &
    \rightarrow
    \begin{cases}
        \makebox[\eqcaselength][l]{%
        $v_1 \texttt{=} v_1 ~ \andoperator$}                                                & \text{if } x \text{ and } y \text{ each comprises}    \\
        ~~ \ldots \andoperator ~ v_n \texttt{=} v_n                                         & ~~ \text{elements } v_1 \ldots v_n                    \\
        \texttt{false}                                                                      & \text{otherwise}                                      \\
    \end{cases}             \\ & \\
f_{\texttt{tuple}}(x, y)    &
    \rightarrow
    \begin{cases}
        \makebox[\eqcaselength][l]{%
        $x\texttt{.}a_1 \texttt{=} y\texttt{.}a_1 ~ \andoperator $}                         & \text{if } x \text{ and } y \text{ each comprises}    \\
        ~~ \ldots \andoperator ~ x\texttt{.}a_n \texttt{=} y\texttt{.}a_n                   & ~~ \text{attributes } a_1 \ldots a_n                  \\
        \texttt{false}                                                                      & \text{otherwise}                                      \\
    \end{cases}             \\ & \\
p \andoperator q            &
    \rightarrow
    \begin{cases}
        \makebox[\eqcaselength][l]{%
        $\texttt{false}$}                   & \text{if } p \text{ or } q \text{ is } \texttt{false}         \\
        \texttt{true}                       & \text{if } p \text{ and } q \text{ are } \texttt{true}        \\
        \texttt{@eq.null\_and\_true}        & \text{if } p \text{ (resp. } q \text{) is } \texttt{null}     \\
                                            & \wedge ~ q \text{ (resp. } p \text{) is } \texttt{true}       \\
        \texttt{@eq.null\_and\_null}        & \text{if } p \text{ and } q \text{ are } \texttt{null}        \\
        \texttt{@eq.null\_and\_missing}     & \text{if } p \text{ (resp. } q \text{) is } \texttt{null}     \\
                                            & \wedge ~ q \text{ (resp. } p \text{) is } \texttt{missing}    \\
        \texttt{@eq.missing\_and\_true}     & \text{if } p \text{ (resp. } q \text{) is } \texttt{missing}  \\
                                            & \wedge ~ q \text{ (resp. } p \text{) is } \texttt{true}       \\
        \texttt{@eq.missing\_and\_missing}  & \text{if } p \text{ and } q \text{ are } \texttt{missing}     \\
    \end{cases}             \\ & \\
\end{array}
\]
\normalsize
\end{mdframed}
\caption{Equality Evaluation Function}
\label{figure:equality:semantics}
\end{figure}

Selection and join queries prominently use the \texttt{=} equality function. Since SQL++ variables bind to arbitrary heterogeneous values and missing information is prevalent, the evaluation function for the equality comparison $x \texttt{=} y$ must specify the semantics of comparing specified (i.e., non-null, non-missing), \texttt{null} and \texttt{missing} values. Furthermore, the lack of schema may lead to comparing mismatched types. For instance, a type mismatch can evaluate to \texttt{false}/\texttt{null}/\texttt{missing} or an error. A language designer can also choose whether to follow SQL's 3-valued logic in which $x$\texttt{=null}$\rightarrow$\texttt{null}.   Analogous decisions apply for \texttt{missing}. Moreover, if deep equality is supported between complex values, there are further options for combining the results of shallow equality, which go beyond \texttt{true}/\texttt{false} to include \texttt{null} and \texttt{missing}.  

    Figure~\ref{figure:equality:bnf} shows the parameters and options of the equality parameter group, which are utilized in the equality evaluation function of Figure~\ref{figure:equality:semantics}. For deep equality comparisons between complex values in $f_{\texttt{array}}$, $f_{\texttt{bag}}$ and $f_{\texttt{tuple}}$, additional subtleties are introduced by \texttt{null} and \texttt{missing} elements/attributes. Consider $f_{\texttt{array}}$ using two arrays $x$ and $y$ that each has $n$ elements, and all elements are either scalar or complex values. Then, each pairwise comparison $x\texttt{[}i\texttt{]} \texttt{=} y\texttt{[}i\texttt{]}$ evaluates to \texttt{true}/\texttt{false}, and array equality is simply the result of the operator $\andoperator$, which returns the conjunction of these \texttt{true}/\texttt{false} values. However, $x\texttt{[}i\texttt{]} \texttt{=} y\texttt{[}i\texttt{]}$ can also evaluate to \texttt{null}/\texttt{missing}.%
$\andoperator$ must be configured to also produce a result for the logical conjunction between \texttt{null}/\texttt{missing}/\texttt{true}/\texttt{false}. This configuration is achieved by the five parameters in Figure~\ref{figure:equality:bnf}, lines~9 to~13, which extend SQL's 3-valued logic.

\begin{figure}
\includegraphics[width=\columnwidth]{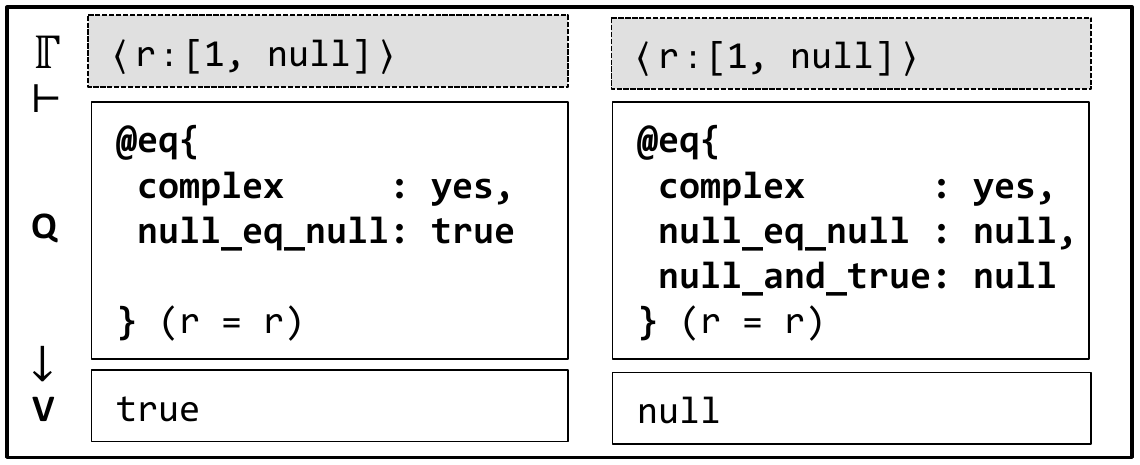}
\caption{Configuring Equality}
\label{figure:equality:example-equality}
\end{figure}

Figure~\ref{figure:equality:example-equality} shows two different configurations and the different results that the same query \texttt{r = r} has for each configuration. The left example follows languages (such as Javascript) which consider \texttt{null} a {\em sentinel} value that is equal to itself. \texttt{null = null} evaluates to \texttt{true}, thus the array \texttt{r} deep equals itself. The right example follows SQL's 3-valued logic, which uses \texttt{null} to represent an unknown value. Since the array contains an unknown value, the result of deep equality is also unknown. Therefore, to support equality that behaves reasonably with respect to \texttt{null}, we recommend language designers utilize the macro config annotation \gl{@unknown.value}. In the case of databases that support \texttt{null} and do not support \texttt{missing}, the two sensible options are \texttt{sentinel} and \texttt{logic}. If a language designer chooses the \texttt{3-valued}, she has implicitly chosen that \texttt{null=null} is \texttt{null} but also that $\texttt{null} o_{and} \texttt{true}$ is \texttt{null}.

\begin{table}[t!]
\centering
\scriptsize
\begin{tabular}{|l|l||l||}
\hline
    \gl{@unknown.value~~~~~~~~:}    & \gl{logic}         & \gl{sentinel}         \\ \hline \hline
\texttt{@eq.null\_eq\_null~~~~~~~:} & \texttt{null}     & \texttt{true}     \\
\texttt{@eq.null\_eq\_missing~~~~:} & \texttt{missing}     & \texttt{false}    \\
\texttt{@eq.null\_eq\_value~~~~~~:} & \texttt{null}     & \texttt{false}    \\
\texttt{@eq.missing\_eq\_missing~:} & \texttt{missing}     & \texttt{true}     \\
\texttt{@eq.missing\_eq\_value~~~:} & \texttt{missing}    & \texttt{false}    \\
\texttt{@eq.null\_and\_true~~~~~~:} & \texttt{null}     &                   \\
\texttt{@eq.null\_and\_null~~~~~~:} & \texttt{null}     &                    \\
\texttt{@eq.null\_and\_missing~~~:} &  \texttt{missing}  &                   \\
\texttt{@eq.missing\_and\_true~~~:} & \texttt{missing} &                   \\
\texttt{@eq.missing\_and\_missing:} &  \texttt{missing} &                  \\
\hline
\end{tabular}
\caption{\gl{@unknown.value} Macro Expansion}
\label{table:equality:macro}
\end{table}

In the presence of \texttt{missing}, the \gl{@unknown.value} must specify its behavior in comparison. In the sentinel approach, the \texttt{missing} will behave as yet another sentinel value. In the logic-based approach, we recall that \texttt{missing} is introduced as ``inapplicable", while \texttt{null} stands for the currently unknown. In such case, the \texttt{logic} approach is expanded as shown in Table~\ref{table:equality:macro}. Notice that Table~\ref{table:equality:macro} provides a pure sentinel approach and an unknown/inapplicable logic approach. One may still consider other possibilities and this is why macro operate as guidelines, rather than definitive options.

\section{SFW Query Clauses}
\label{section:clauses}

\subsection{\texttt{FROM} clause}
\label{section:from}

\begin{figure}
\scriptsize
\begin{tabular}{|@{~}rc@{~}l@{~}|}
\hline
\hline
         1  & \multicolumn{2}{@{}l@{~}|}{\gn{from\_clause}} \\
         2  & \gp   & \gl{FROM} \gn{from\_item} \\ \hline
         3  & \multicolumn{2}{@{}l@{~}|}{\gn{from\_item}} \\
         4  & \gp   & \gn{expr\_query} \gl{AS} \gn{var} (\gl{AT} \gn{var})? \\
         5  & \gd   & \gn{expr\_query} \gl{AS} \gl{\{} \gn{var} \gl{:} \gn{var} \gl{\}} \\
         6  & \gd   & \gn{from\_item} (\gl{INNER}\gd\gl{LEFT} \gl{OUTER}?) \gl{CORRELATE}? \gn{from\_item} \\
         7  & \gd   & \gn{from\_item} \gl{FULL} \gl{OUTER}? \gl{CORRELATE}? \gn{from\_item} \gl{ON} \gn{expr\_query} \\
         8  & \gd   & \gn{from\_item} \gl{,} \gn{from\_item} \\
         9  & \gd   & \gn{from\_item} (\gl{INNER}\gd\gl{LEFT}\gd\gl{RIGHT}\gd\gl{FULL}) \gl{JOIN} \gn{from\_item} \\
        10  &       & ~~~~ \gl{ON} \gn{expr\_query} \\
        11  & \gd   & (\gl{INNER}\gd\gl{OUTER}) \gl{FLATTEN(}\gn{expr\_query} \gl{AS} \gs{var}\gl{,} \\
        12  &       & ~~~~\gn{expr\_query} \gl{AS} \gs{var}\gl{)} \\ \hline
        13  & \multicolumn{2}{@{}l@{~}|}{\gn{from\_param}} \\
        14  & \gp   & \gl{bag\_order:}  (\gl{counter}\gd\gl{null}\gd\gl{missing}\gd\gl{error}) \\
        15  & \gd   & \gl{coerce\_null\_to\_collection~~~:} (\gl{singleton}|\gl{empty}|\gl{error}) \\
        16  & \gd   & \gl{coerce\_missing\_to\_collection:} (\gl{singleton}|\gl{empty}|\gl{error}) \\
        17  & \gd   & \gl{coerce\_value\_to\_collection~~:} (\gl{singleton}|\gl{error}) \\
        18  & \gd   & \gl{coerce\_null\_to\_tuple~~~:} (\gl{empty}|\gl{error}) \\   
        19  & \gd   & \gl{coerce\_missing\_to\_tuple:} (\gl{empty}|\gl{error}) \\
        20  & \gd   & \gl{no\_match:}  (\gl{null}\gd\gl{missing}) \\

\hline
\hline
\end{tabular} 
\caption{BNF Grammar for \texttt{FROM} Clause and \texttt{@from} Config Parameters}
\label{figure:from:bnf}
\end{figure}

\newcommand{\linefrom}[1]{%
    \IfEqCase*{#1}{%
    {single}{Lines~4-5}%
    {correlate}{Lines~6-7}%
    {sugar}{Lines~8-12}%
    {collection}{(line~4)}%
    {bag order}{(Figure~\ref{figure:from:bnf}, line~14)}%
    {coerce collection}{(Figure~\ref{figure:from:bnf}, lines 15-17)}%
    {coerce tuple}{(Figure~\ref{figure:from:bnf}, lines 18-19)}%
    {tuple}{(Figure~\ref{figure:from:bnf}, line~5)}%
    {inner correlate}{(Figure~\ref{figure:from:bnf}, line~6)}%
    {left correlate}{(Figure~\ref{figure:from:bnf}, line~6)}%
    {no match}{(line~20)}%
    {full correlate}{(Figure~\ref{figure:from:bnf}, line~7)}%
    {cartesian product}{(Figure~\ref{figure:from:bnf}, line~8)}%
    {join}{(lines~9-10)}%
    {flatten}{(lines~11-12)}%
    }[\errmessage{Unable to ref #1 for FROM BNF}]%
}%

The SQL++ \texttt{FROM} clause allows {\em variables} to range over any data, unlike SQL \texttt{FROM} clause tuple variables that range over tuples only. The SQL++  variables can range over heterogeneous elements, over nested collections, and over the attribute/value pairs of tuples (a feature reminiscent of SQL extensions for dynamic schema introspection \cite{schemasql-vldb-1996}). Furthermore, \texttt{FROM} variables can register the order of input elements in arrays, which is useful for order-aware queries.

Some semi-structured query languages have introduced {\em unnesting} constructs (following the nested relational algebra \cite{nest-unnest-pods-1982,nested-relational-vldb-1988}) that essentially specify how the variables of the \texttt{FROM} clause range over nested data. The SQL++ core semantics show that these constructs can be fused with SQL and be expressed succinctly by the combination of (a) allowing correlated queries in SQL's \texttt{FROM} clause, and (b) a distinction of how variable bindings of the correlated queries combine, which is akin to the distinction between \texttt{JOIN} versus \texttt{LEFT OUTER JOIN} in SQL. The combination of these two simple features, allows the SQL++ core to directly express queries that could only be expressed in cumbersome ways (involving intermediate results) even in advanced prior languages, such as XQuery and the nested relational algebra.

Figure~\ref{figure:from:bnf} shows the BNF grammar for the SQL++ \texttt{FROM} clause and corresponding config parameters. The syntax and semantics of \texttt{FROM} are defined inductively with three cases. (1) \linefrom{single}: the SQL++ core base case where the \texttt{FROM} item ranges over a single collection or tuple (2) \linefrom{correlate}: the SQL++ core inductive case where the \texttt{FROM} item comprises correlation between two other \texttt{FROM} items (3) \linefrom{sugar}: the ``syntactic sugar'' cases, where the grammar introduces well known SQL constructs (e.g., joins, outer joins) as well as the unnesting constructs of NoSQL databases. The semantics of the ``syntactic sugar" constructs are explained by reduction to the SQL++ core constructs of \#1 and \#2.

\begin{figure}
\includegraphics[width=\columnwidth]{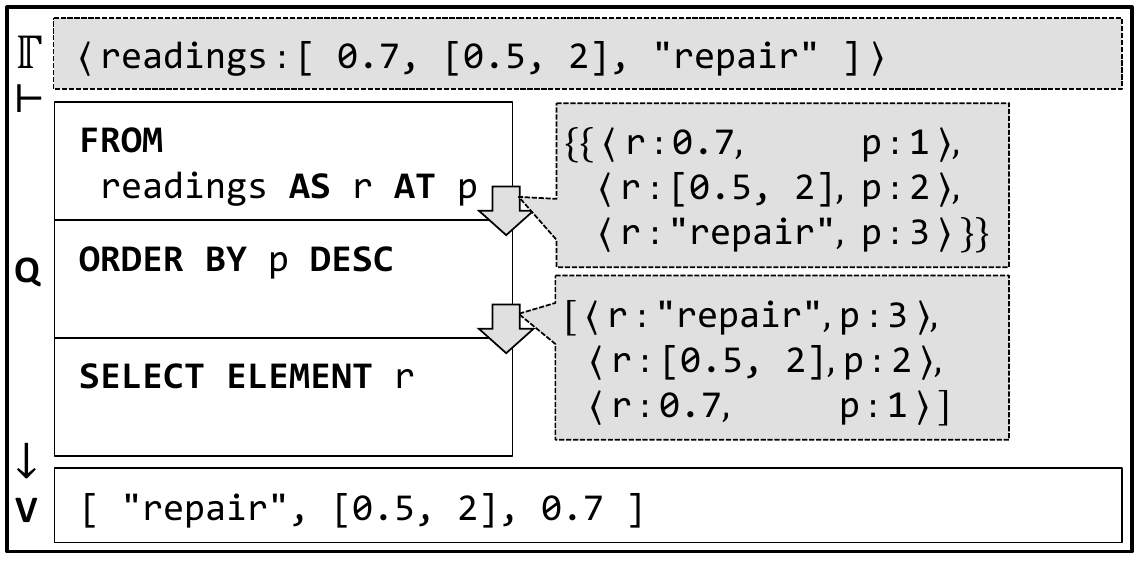}
\caption{Using \texttt{AT} (Position Variable) to Obtain Ordinal Positions for Input Order}
\label{figure:from:example-from-collection}
\end{figure}

\highlight{(1a) SQL++ Core, Ranging over Collection Elements} \texttt{FROM } $e$ \texttt{AS} $x$ (\texttt{AT } $y$)? supports ranging over elements of a collection \linefrom{collection}. 
$x$ is an \textit{element variable}, and $y$ is an optional \textit{position variable}. As in SQL, expression $e$ can be any SFW subquery or named value \linequery{expression, sql}. SQL++ extends beyond SQL towards full composability, as $e$ can also be a variable, path, function call, value constructor or value literal \linequery{expression, extensions}.

Let $e \rightarrow c$, where $c$ is a collection
with $n$ elements $u_1, \ldots, u_n$. For each $u_j, ~ j=1\ldots,n$, the \texttt{FROM} clause outputs a binding tuple $\langle x : u_j, ~ y : v_j \rangle$, where $v_j$ is the ordinal position of $u_j$ in $c$ when $c$ is an array, or a value specified by the config parameter \texttt{@from.bag\_order} \linefrom{bag order} when $c$ is a bag. A language designer sets \texttt{@from.bag\_order} to \texttt{counter} to produce a random order and assign sequential numbers accordingly. Alternately, uses \texttt{null} or \texttt{missing} as sentinel values indicating bag positions are meaningless.
As in SQL, the SQL++ \texttt{FROM} clause produces an unordered bag of binding tuples. An output order can be later imposed by \texttt{ORDER BY}, which may utilize the position variable $y$.

For example, Figure~\ref{figure:from:example-from-collection} shows a SQL++ query that reverses an array. The \texttt{FROM} clause defines element variable \texttt{r} and position variable \texttt{p}, and outputs 3 binding tuples that respectively bind \texttt{r} to elements and \texttt{p} to positions. The \texttt{ORDER BY} clause uses \texttt{p} to sort the binding tuples in descending order of their original positions in the array, and the \texttt{SELECT ELEMENT} clause outputs the reversed array elements.

Since SQL++ is semi-structured, it is possible that $e$ evaluates into a value $c$ that is not a collection. In such case, the language designer may choose to coerce $c$ into a collection, according to one of the ``coerce'' configuration options \linefrom{coerce collection}. For example, if \texttt{@from.coerce\_null\_to\_collection} is set to \texttt{singleton}, then when $c$ is \texttt{null}, $c$ is coerced into \texttt{\{\!\{null\}\!\}}. If set to \texttt{empty}, then \texttt{null} is coerced into the empty collection \texttt{\{\!\{\}\!\}}. If $c$ is a (scalar or tuple) value then the designer may choose to coerce it into the collection $\texttt{\{\!\{}c\texttt{\}\!\}}$, by setting \texttt{@from.coerce\_value\_to\_collection} to \texttt{singleton}.

\begin{figure}
\includegraphics[width=\columnwidth]{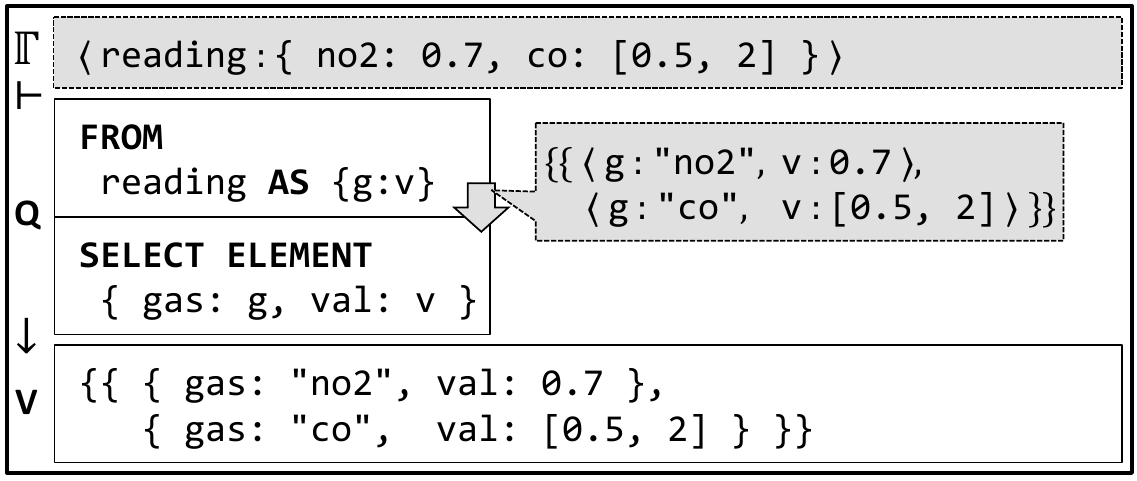}
\caption{Using \texttt{AS \{x:y\}} to Range Over Tuple Attributes}
\label{figure:from:example-from-tuple}
\end{figure}

\highlight{(1b) SQL++ Core, Ranging over Tuple Attributes} \texttt{FROM} $t$ \texttt{AS \{}$a$\texttt{:}$v$\texttt{\}} enables ranging over the attribute name-value pairs of a tuple \linefrom{tuple}. 
When expression $t$ evaluates to a tuple $\{a_1:v_1, \ldots,a_n:v_n\}$, \texttt{FROM} outputs the bag of binding tuples $\{\!\{ \langle a:a_1, v:v_1\rangle \ldots \langle a:a_n, v:v_n\rangle \}\!\}$.

Figure~\ref{figure:from:example-from-tuple} shows an example SQL++ query that inputs a tuple with 2 attributes \texttt{no2} and \texttt{co}. The \texttt{FROM} clause defines variables $g$ and $v$, and outputs 2 binding tuples where $g$ is bound to the attribute name and $v$ is bound to the attribute value. The \texttt{SELECT ELEMENT} clause outputs a bag with 2 elements, one for each input tuple attribute. In SQL's parlance, the SQL++ query has transposed/pivoted the tuple's attributes (columns) into the collection's elements (rows).

If $t$ evaluates into a value that is not a tuple, coercion may be activated by the configuration options of \linefrom{coerce tuple}: When $t$ evaluates to \texttt{null} or \texttt{missing}, the language designer may elect to coerce $t$ into the empty tuple \texttt{\{\}}, instead of producing an error.

\begin{figure}
\includegraphics[width=\columnwidth]{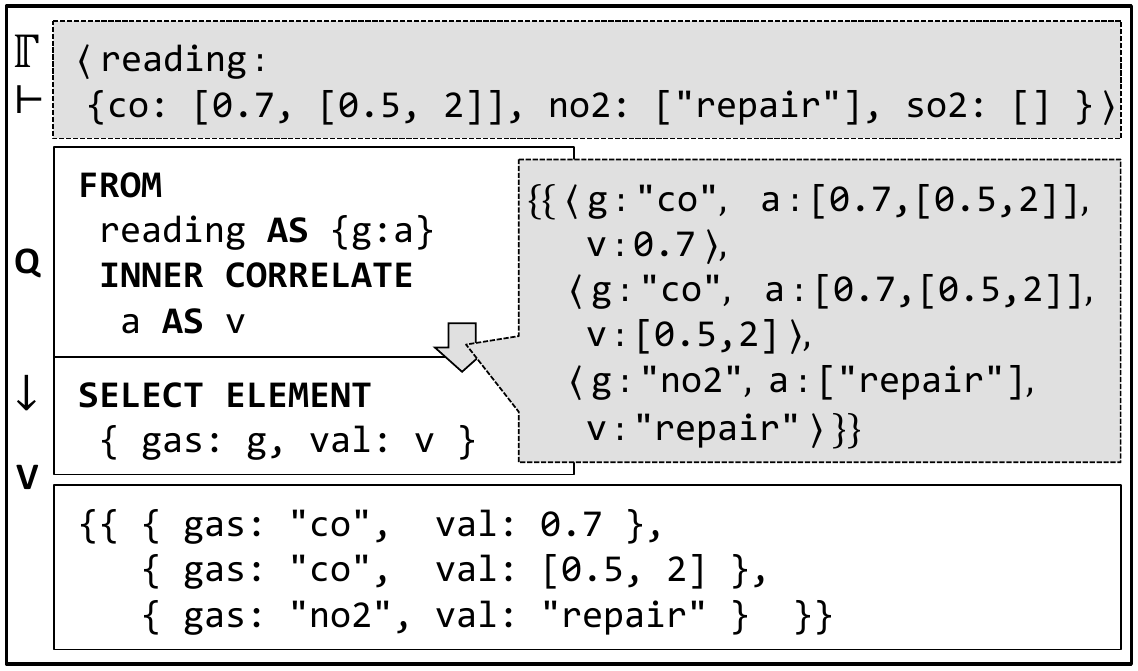}
\caption{Using \texttt{INNER CORRELATE} to Correlate \texttt{FROM} Items}
\label{figure:from:example-from-correlate}
\end{figure}

\highlight{(2a) Core, \texttt{INNER CORRELATE}} The first inductive step case is $\texttt{FROM } l \texttt{ INNER CORRELATE } r$ \linefrom{inner correlate}. (For brevity, also written as \texttt{FROM} $l$ \texttt{INNER} $r$.) The \texttt{FROM} item $r$ can utilize variables defined by \texttt{FROM} item $l$. Suppose the SFW query is evaluated in binding env $\env$. Let $B^l$ be the bag of binding tuples that would result from the clause $\texttt{FROM } l$. Then, $\texttt{FROM } l \texttt{ INNER CORRELATE } r$ outputs all binding tuples $b^l_i \| b^r_{i,j}$, where $b^l_i \in B^l$, $b^r_{i,j} \in B^r_i$, and $B^r_i$ is the bag of binding tuples that would result from evaluating $\texttt{FROM } r$ within the binding env $b^l_i \| \env$.

For example, Figure~\ref{figure:from:example-from-correlate} shows a SQL++ query that ranges over attribute/value pairs, thereafter ranges over the array of each attribute value. \texttt{FROM reading AS \{g:a\}} would result in binding tuples $b^l_1$, $b^l_2$, $b^l_3$, each with variables \texttt{g} and \texttt{a}. Evaluating the right hand side \texttt{a AS v} in the context of $b^l_1 = \langle \texttt{g} : \texttt{"co"}, \texttt{a} : \texttt{[0.7,[0.5,2]]} \rangle$ results into $B^r_1 = \{\!\{ \langle \texttt{v} : \texttt{0.7} \rangle, \langle \texttt{v} : \texttt{[0.5,2]} \rangle \}\!\}$. Consequently the \texttt{FROM} clause outputs the first 2 binding tuples of Figure~\ref{figure:from:example-from-correlate}. Likewise $b^l_2 = \langle \texttt{g}:\texttt{no2}, \texttt{a}:\texttt{["repair"]}\rangle$, leads to $B^r_2 = \{\!\{\langle \texttt{v}:\texttt{"repair"}\rangle\}\!\}$ which leads to \texttt{FROM} outputting the 3rd binding tuple. Since $b^l_3 = \langle \texttt{g} : \texttt{"so2"}, \texttt{a} : \texttt{[]} \rangle$ results in an empty $B^r_3$, there is no binding tuple that corresponds to \texttt{so2}.

Notice that correlated queries are akin to nested for-loops of imperative languages, where the left subquery $l$ and the right subquery $r$ are respectively the outer and inner loops. While this example has a simple expression $r$ (i.e. the expression \texttt{a AS v}), in general $r$ can be a complex SFW query. For example, we could replace the \texttt{a AS v} in Figure~\ref{figure:from:example-from-correlate} with the following subquery, which would correlate each \texttt{g}/\texttt{a} pair only with the top-1 element of each collection.\\
\texttt{(FROM ~~~~a AS x} \\
\texttt{\phantom{ }ORDER BY x} \\
\texttt{\phantom{ }LIMIT ~~~1} \\
\texttt{\phantom{ }SELECT ~~ELEMENT x) AS v}

\highlight{(2b) Core, \texttt{LEFT OUTER CORRELATE}} In \texttt{FROM} $l$ \texttt{LEFT} \texttt{OUTER} \texttt{CORRELATE} $r$ \linefrom{left correlate}, $r$ can again utilize variables defined by $l$. (For brevity one may simply write $l$ \texttt{LEFT} $r$, omitting either or both of the \texttt{OUTER} and \texttt{CORRELATE} keywords.) Similar to SQL's \texttt{LEFT JOIN}, the \texttt{LEFT} outputs binding tuples derived from $l$, even if there are no matching binding tuples from $r$. Formally, let $B^l$ be the bag of binding tuples that results from the clause $\texttt{FROM } l$. Then, $\texttt{FROM } l \texttt{ LEFT } r$ outputs all binding tuples that $\texttt{FROM } l \texttt{ INNER } r$ would output. In addition, for each $b^l_i \in B^l$ such that $B^r_i$ is an empty bag, \texttt{LEFT} also outputs a binding tuple $b^l_i \| \langle x^r_1: \texttt{@from.no\_match}, \ldots, x^r_n: \texttt{@from.no\_match} \rangle$, where $x^r_1, \ldots, x^r_n$ are variables defined by $r$, and each variable is bound to the value configured by \texttt{@from.no\_match} \linefrom{no match}. Since \texttt{LEFT} generalizes SQL's \texttt{LEFT JOIN} (as discussed below), achieving SQL compatibility requires configuring \texttt{@from.no\_match} to \texttt{null}. Alternatively, a language designer may also choose configuring to \texttt{missing}. The extended version argues that the latter is preferrable because (i) it distinguishes the kinds of information absence (when the language designer wants \texttt{null} to mean ``unknown" and \texttt{missing} to mean ``inapplicable" and (2) enables simple rewritings that decorrelate $r$ and open the gates to set-at-a-time processing, of the kind accomplished by SQL processors.

For example, suppose Figure~\ref{figure:from:example-from-correlate} uses \texttt{LEFT} instead of \texttt{INNER}. Assuming \texttt{@from.no\_match} is configured to \texttt{null}, the query will output the additional tuple: \texttt{\{gas:"so2", val:null\}}.

\highlight{(2c) Core, \texttt{FULL CORRELATE}} \texttt{FROM} $l$ \texttt{FULL} \texttt{CORRELATE} $r$ \texttt{ON} $c$ \linefrom{full correlate} resembles SQL's \texttt{FULL JOIN}: $l$ and $r$ cannot utilize variables of each other, and their correlation is specified by the condition $c$. Due to its similarity to SQL's \texttt{FULL JOIN}, we omit a formal specification.

Formally, let $B^l$ (resp. $B^r$) be the bag of binding tuples resulting from the clause $\texttt{FROM } l$ (resp. $\texttt{FROM } r$). Also, let $l$ (resp. $r$) define variables $x^l_1, \ldots, x^l_m$ (resp. $y^r_1, \ldots, y^r_n$). Then, $\texttt{FROM } l \texttt{ FULL CORRELATE } r \texttt{ ON } c$ outputs all binding tuples $b^l_i \| b^r_{j}$, where $b^l_i \in B^l$, $b^r_j \in B^r$ and $c \rightarrow \texttt{true}$ within binding env $b^l_i \| b^r_j \| \env_b$. In addition, for each $b^l_i$ where there is no $b^r_j$ that matches (i.e. $c \rightarrow \texttt{true}$), \texttt{FULL CORRELATE} also outputs a binding tuple $b^l_i \| \langle y^r_1: \texttt{@from.no\_match}, \ldots, y^r_n: \texttt{@from.no\_match} \rangle$. Conversely, for each $b^r_j$ where there is no matching $b^l_i$, \texttt{FULL CORRELATE} also outputs a binding tuple $\langle x^l_1: \texttt{@from.no\_match}, \ldots, x^l_m: \texttt{@from.no\_match} \rangle \| b^r_j$.

\highlight{(3a) Syntactic Sugar, SQL's Cartesian Product} For SQL compatibility, SQL++ supports $\texttt{FROM } l \texttt{ , } r$ \linefrom{cartesian product} as a special case of $\texttt{FROM } l \texttt{ INNER } r$, in which $r$ does not utilize variables of $l$.

\highlight{(3b) Syntactic Sugar, SQL's Join} Similarly, SQL++ supports SQL joins \linefrom{join} as a special case of \texttt{CORRELATE}. \texttt{LEFT JOIN} is syntactic sugar for the core \texttt{LEFT CORRELATE} as follows, and analogously, \texttt{INNER JOIN} is syntactic sugar for \texttt{INNER CORRELATE}. Finally, \texttt{RIGHT JOIN} is syntactic sugar for \texttt{LEFT JOIN}, and \texttt{FULL JOIN} is identical to \texttt{FULL CORRELATE}. \textbf{(S)} denotes syntactic sugar and \textbf{(C)} denotes core SQL++.  
~\\
\begin{tabular}{@{}l@{~}l@{~}l@{~}l@{}}
\textbf{(S)}    & \texttt{FROM} $l$ \texttt{AS} $x$                         & \textbf{(C)}  & \texttt{FROM} $l$ \texttt{AS} $x$ \\
                & ~\texttt{LEFT JOIN} $r$ \texttt{AS} $y$ \texttt{ON} $c$   &               & ~\texttt{LEFT CORRELATE (} \\
                &                                                           &               & ~~\texttt{FROM~~} $r$ \texttt{AS} $y$ \\
                &                                                           &               & ~~\texttt{WHERE~} $c$ \\
                &                                                           &               & ~~\texttt{SELECT} \texttt{ELEMENT} $y$ \texttt{) AS} $y$ \\
\end{tabular} \\

\highlight{(3c) Syntactic Sugar, Flattening} Flattening (aka unnesting) is commonly used within SQL-on-Hadoop and NoSQL databases to range over nested collections. SQL++ supports \texttt{INNER FLATTEN} \linefrom{flatten} as a special case of the core \texttt{INNER}, in which $r$ must be a path expression $x.p$ that starts from the left hand side variable $x$, as follows. Analogously, \texttt{OUTER FLATTEN} is syntactic sugar for \texttt{LEFT}.
~\\
\begin{tabular}{@{}l@{~}l@{~}l@{~}l@{}}
\textbf{(S)}    & \texttt{FROM INNER FLATTEN(}                                      & \textbf{(C)}  & \texttt{FROM} $l$ \texttt{AS} $x$ \\
                & ~$l$ \texttt{AS} $x$\texttt{,} $x.p$ \texttt{AS} $y$\texttt{)}    &               & ~\texttt{INNER CORRELATE} $x.p$ \texttt{AS} $y$ \\
\end{tabular} \\

\subsection{\texttt{SELECT} clause}
\label{section:select}

\begin{figure}
\scriptsize
\begin{tabular}{|@{~}rc@{~}l@{~}|}
\hline
\hline
         1  & \multicolumn{2}{@{}l@{~}|}{\gn{select\_clause}} \\
         2  & \gp   & \gl{SELECT ELEMENT} \gn{expr\_query} \\
         3  & \gp   & \gl{SELECT ATTRIBUTE} \gn{expr\_query} \gl{:} \gn{expr\_query} \\
         4  & \gp   & \gl{SELECT} \gn{expr\_query} (\gl{AS} \gn{attr\_name})? \\
         5  &       & ~~~~(\gl{,} \gn{expr\_query} (\gl{AS} \gn{attr\_name})?)* \\
\hline
\hline
\end{tabular} 
\caption{BNF Grammar for \texttt{SELECT} Clause}
\label{figure:select:bnf}
\end{figure}

\newcommand{\lineselect}[1]{%
    \IfEqCase*{#1}{%
    {element}{(Figure~\ref{figure:select:bnf}, line~2)}%
    {attribute}{(Figure~\ref{figure:select:bnf}, line~3)}%
    {nested}{(Figure~\ref{figure:select:bnf}, lines~2-3)}%
    {sugar}{(Figure~\ref{figure:select:bnf}, lines~4-5)}%
    }[\errmessage{Unable to ref #1 for SELECT BNF}]%
}%

The \texttt{SELECT} clause of SQL++ can output more than SQL's collections of flat tuples. First, an output tuple can contain other nested collections. Second, an output collection can directly contain (i.e. without intervening tuples) any arbitrary value, including scalars, collections, \texttt{null} and \texttt{missing}. Finally, output tuples can be heterogeneous: Missing attributes in the input can reflect to missing attributes in the output. Indeed, an output tuple can have a data-dependent number of attributes, where both the attribute names and their values are provided by variables. 

\highlight{(1) SQL++ Core, Outputting a Collection} \texttt{SELECT ELEMENT} $e$ \lineselect{element} is core SQL++ that outputs a collection of arbitrary elements (both tuples and non-tuples). Notice that the keywords are different from SQL's plain \texttt{SELECT}, which is subsequently explained as syntactic sugar for outputting specifically a collection of tuples. The \texttt{SELECT ELEMENT} clause inputs a bag (resp. array, if \texttt{ORDER BY} is present) of binding tuples $B^{in}_{\texttt{SELECT}}$, and outputs a bag (resp. array) of values. Let $\env$ be the binding env of the enclosing SFW query. For each input binding tuple $b \in B^{in}_{\texttt{SELECT}}$, \texttt{SELECT ELEMENT} outputs a value $v$, where $b \| \env \vdash e \rightarrow v$. Typically $e$ is a tuple/array/bag constructor \linequery{constructors}, but in general $e$ can be any expression. Figures~\ref{figure:sfw:example-sfw}, \ref{figure:from:example-from-collection} and \ref{figure:from:example-from-tuple} show respective examples of a path expression, a plain variable and a tuple constructor within \texttt{SELECT ELEMENT}.

\begin{figure}
\includegraphics[width=\columnwidth]{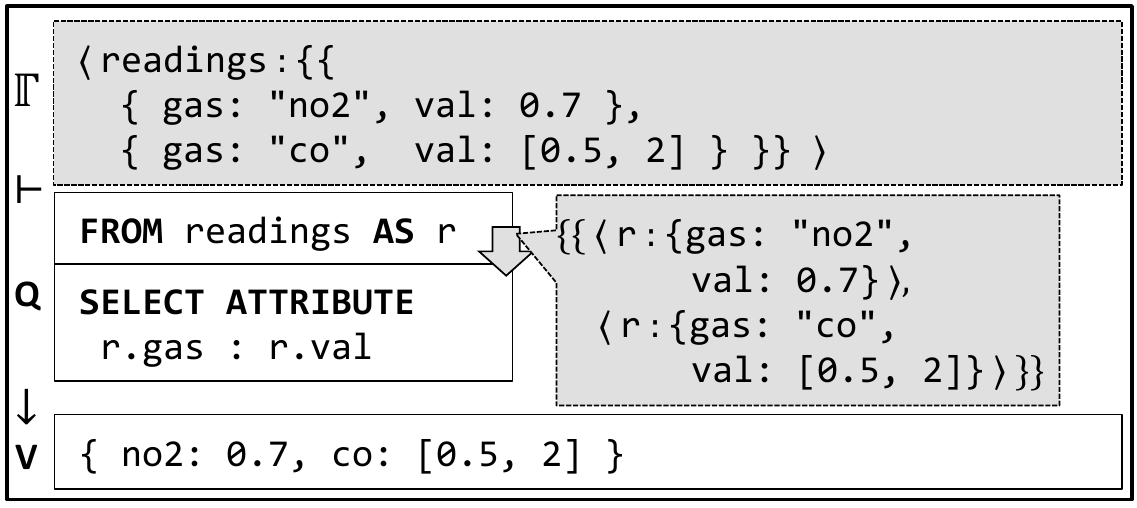}
\caption{Using \texttt{SELECT ATTRIBUTE} to Output a Variable-Width Tuple}
\label{figure:select:example-select-attribute}
\end{figure}

When a \texttt{missing} is provided as an attribute value in a constructor then the particular attribute/value pair is omitted. Therefore, particular attribute omissions and heterogeneities in the input can easily propagate to corresponding attribute omissions and heterogeneities in the output. For example, consider the query 
\texttt{FROM readings AS r SELECT ELEMENT \{"co":r.co, "no2":r.no2\}} 
evaluated on the \texttt{readings} heterogeneous collection of tuples of Figure~\ref{figure:sfw:example-sfw}. The query output is identical to \texttt{readings}, since the absence of the attribute \texttt{no2} in the first tuple of the input will lead to a respective absence of \texttt{no2} in the first tuple of the output.

\highlight{(2) SQL++ Core, Outputting a Variable-Width Tuple} \texttt{SELECT ATTRIBUTE} $e_n$\texttt{:}$e_v$ \lineselect{attribute} outputs a tuple where the attribute name-value pairs are determined by the input bindings. For each input binding tuple $b$ of $B^{in}_{\texttt{SELECT}}$, the output tuple contains an attribute $n : v$, where $b \| \env \vdash e_n \rightarrow n$ and $b \| \env \vdash e_v \rightarrow v$. An error occurs if $n$ is not a string.  The attribute/value pair is omitted if $v$ is \texttt{missing}.

Figure~\ref{figure:select:example-select-attribute} shows a collection-to-tuple pivoting example. The query output tuple contains two attributes whose names correspond to the evaluation of \texttt{r.gas} in the first binding tuple (\texttt{no2}) and the second binding tuple (\texttt{co}). Notice that Figure~\ref{figure:select:example-select-attribute}'s query is the inverse of Figure~\ref{figure:from:example-from-tuple}'s, and supports the complementary transposing/pivoting of the input collection's elements (rows) into the output tuple's attributes (columns).

\begin{figure}
\includegraphics[width=\columnwidth]{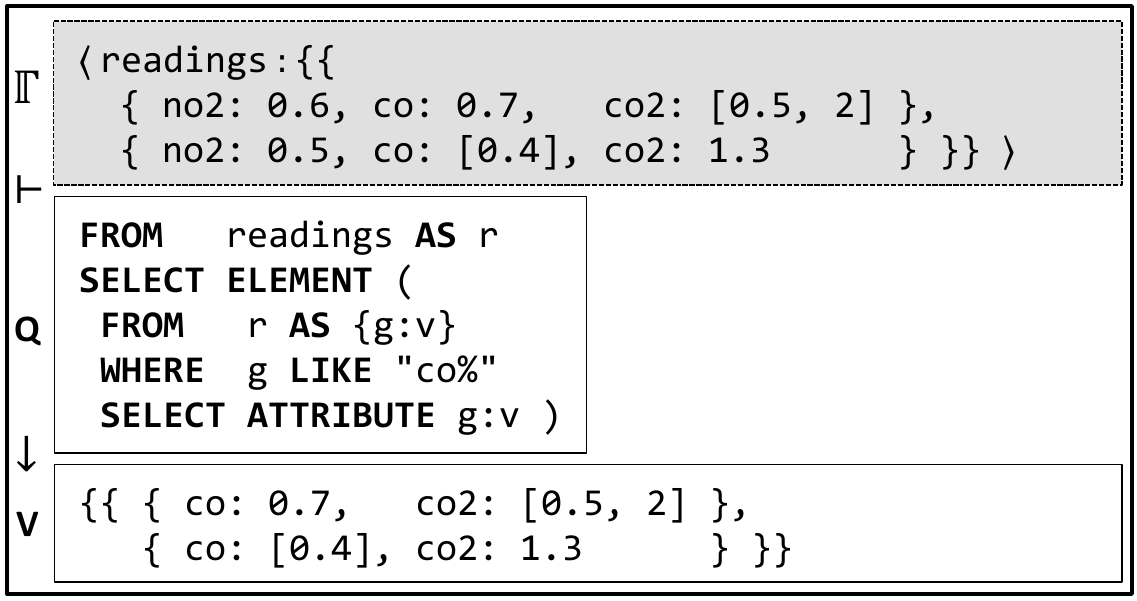}
\caption{\texttt{SELECT ATTRIBUTE} nested in \texttt{SELECT ELEMENT}}
\label{figure:select:example-select-nested}
\end{figure}

\highlight{(3) Core, Outputting Nested Values} Both \texttt{SELECT ELEMENT} and \texttt{SELECT ATTRIBUTE} can comprise nested SFW queries \lineselect{nested}. For example, nesting a SFW query with \texttt{SELECT ELEMENT} within a parent \texttt{SELECT ELEMENT} clause will output nested collections. Figure~\ref{figure:select:example-select-nested} shows another example, where \texttt{SELECT ATTRIBUTE} is  nested within a parent \texttt{SELECT ELEMENT} clause and outputs a collection of variable-width tuples. For each tuple in the input collection, the query outputs a tuple that contains only attributes that start with \texttt{co}.    

\highlight{(4) Syntactic Sugar, SQL's \texttt{SELECT}} For SQL compatibility, SQL++ supports \texttt{SELECT} $e_1$ \texttt{AS} $a_1$ \texttt{,} $\ldots$ \texttt{,} $e_n$ \texttt{AS} $a_n$ \lineselect{sugar} as syntactic sugar for \texttt{SELECT ELEMENT \{}$a_1$\texttt{:}$e_1$ \texttt{,} \ldots \texttt{,} $a_n$\texttt{:}$e_n$\texttt{\}}, i.e., a query that outputs a collection of tuples constructed with attributes $a_1, \ldots, a_n$. Furthermore, when expression $e_i$ is of the form $e'\texttt{.}n$ (i.e. a path that navigates into tuple attribute $n$), SQL++ follows SQL in allowing \texttt{AS} $a_i$ (i.e. attribute names) to be optional. In this case, \texttt{SELECT} $e'\texttt{.}n$ is equivalent to \texttt{SELECT} $e'\texttt{.}n$ \texttt{AS} $n$.

\subsection{\texttt{GROUP BY} clause}
\label{section:group-by}

\begin{figure}
\includegraphics[width=\columnwidth]{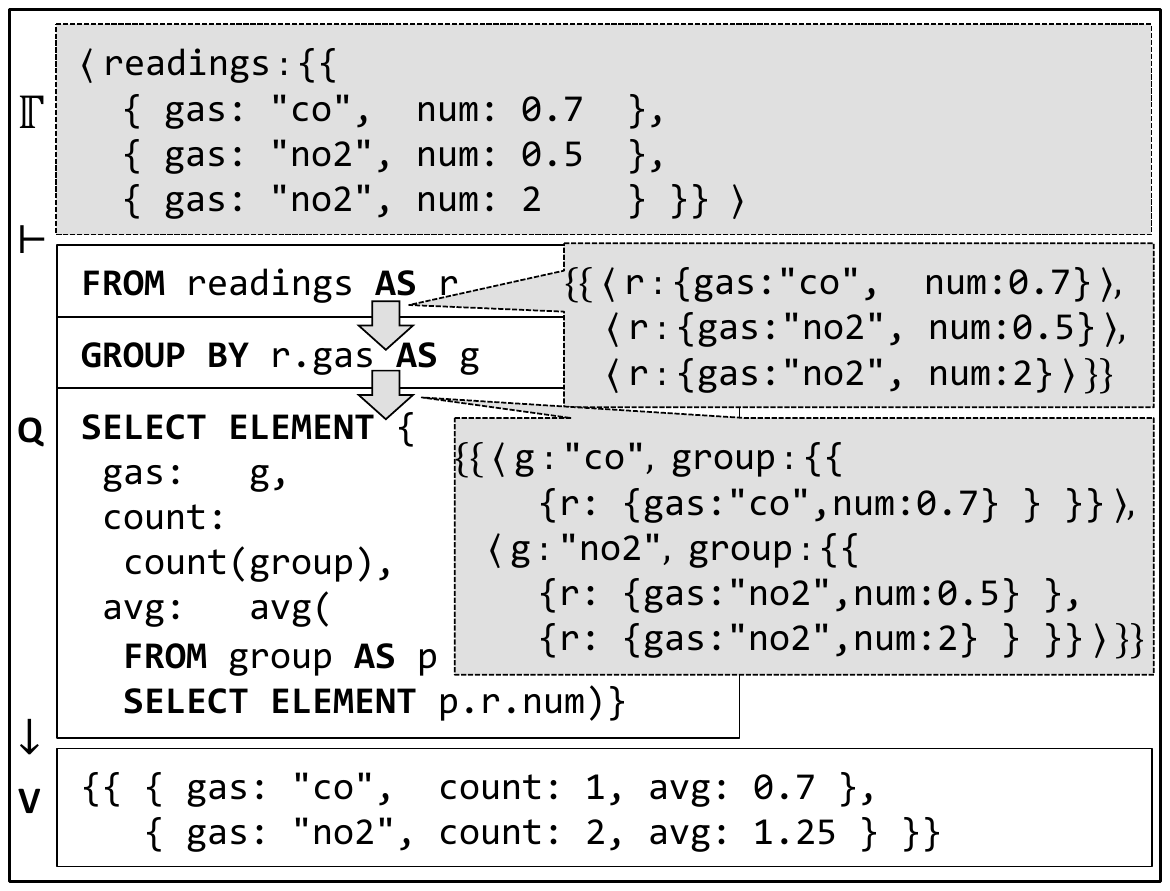}
\caption{Using \texttt{GROUP BY} to Group Elements}
\label{figure:group-by:example-group-by}
\end{figure}

The SQL++ \texttt{GROUP BY} clause generalizes grouping to unify SQL's semantics for 3 separate classes of functions: (i) ordinary functions (e.g. \texttt{LIKE}) that input scalars and output scalars (ii) aggregation functions (e.g. \texttt{SUM}) used with \texttt{GROUP BY} to input bags and output scalars (iii) window functions (e.g. \texttt{ROW\_NUMBER}) used with \texttt{OVER} / \texttt{PARTITION BY} to input bags and output bags.

\highlight{(1) Core, \texttt{group} Variable} \texttt{GROUP BY} $e_1$ \texttt{AS} $x_1$\texttt{,} \ldots \texttt{,} $e_m$ \texttt{AS} $x_m$ \linequery{group by} is core SQL++ that creates groups. Each $e_i$ is a \textit{grouping expression}, and each $x_i$ is a \textit{grouping variable}. Let $\env_b$ be the binding env of the enclosing SFW query. The bag of input binding tuples $B^{in}_{\texttt{GROUP}}$ are partitioned into the minimal number of equivalence groups $B_1 \ldots B_n$, such that any two binding tuples $b, b' \in B^{in}_{\texttt{GROUP}}$ are in the same equivalence group if and only if every grouping expression $e_i$ evaluates to the identical value $v_i$,  i.e., $b || \env_b \vdash e_i \mapsto v_i$ and $b' || \env_b \vdash e_i \mapsto v_i$.
For each group $B_j ~ (1 \leq j \leq n)$, the \texttt{GROUP BY} clause outputs a binding tuple $b_j =\langle x_1 : v_1, ~ \ldots, ~ x_m : v_m, ~ \texttt{group} : B_j \rangle$. 

Notice that unlike SQL, the output binding tuple provides the partitioned input binding tuples in the special variable \texttt{group}, which can be explicitly utilized in subsequent \texttt{HAVING}, \texttt{ORDER BY} and \texttt{SELECT} clauses. Thus, a SQL++ query can perform complex computations on the groups, leading to results of any type (e.g. collections nested within collections). The explicit presence of groups in SQL++, while more general than SQL, also leads to simpler semantics than those of SQL, since the \texttt{GROUP BY} clause semantics are independent of the presence of subsequent functions in \texttt{HAVING}, \texttt{ORDER BY} and \texttt{SELECT}.

For example, Figure~\ref{figure:group-by:example-group-by} shows a SQL++ query for counting and averaging the readings of each gas. The \texttt{GROUP BY} clause inputs 3 binding tuples, and uses \texttt{r.gas} to partition them into 2 groups. Within \texttt{SELECT ELEMENT}, the \texttt{count} function counts the elements in each \texttt{group}, whereas \texttt{avg} outputs the average \texttt{r.num} in each \texttt{group}. The \texttt{GROUP BY} clause is independent of \texttt{count} and \texttt{avg}, since both are general-purpose functions that can input arbitrary collections (i.e. not just \texttt{group}).

\highlight{(2) Syntactic Sugar, SQL's \texttt{GROUP BY}} For SQL compatibility, SQL++ supports using a grouping expression $e_i$ in \texttt{HAVING}, \texttt{ORDER BY} and \texttt{SELECT} clauses, and omitting its corresponding grouping variable (i.e. \texttt{AS} $x_i$) in the \texttt{GROUP BY} clause. Thus, $e_i$ becomes syntactic sugar for $x_i$ within \texttt{HAVING}, \texttt{ORDER BY} and \texttt{SELECT}. For example, Figure~\ref{figure:group-by:example-group-by} can equivalently use: \texttt{SELECT ELEMENT \{ gas : r.gas} $\ldots$ \texttt{\}}.

\highlight{(3) Syntactic Sugar, SQL's Aggregation Functions} SQL++ also supports syntactic sugar for aggregation functions. Suppose any of the \texttt{SELECT}, \texttt{HAVING} and \texttt{ORDER BY} clauses contains a function call $f(e)$, where $f$ is a SQL aggregation function such as \texttt{sum} and \texttt{avg}, and expression $e$ utilizes variables $v_1, \ldots, v_n$ that are defined in the \texttt{FROM} clause. Then, $f(e)$ is syntactic sugar for $f(\texttt{SELECT}\ e'\ \texttt{FROM\ group AS}\ p)$, where $e'$ results from substituting each $v_i$ in $e$ with $p.v_i$. For the special case where an aggregation function $f$ (such as \texttt{count}) needs the entire \texttt{group} as-is, $f(*)$ is syntactic sugar for $f(\texttt{group})$. For example, Figure~\ref{figure:group-by:example-group-by} can equivalently use: \texttt{count(*)} and \texttt{avg(r.num)}.

\section{SQL++ as an Expressiveness Benchmark of Databases}
\label{section:experiments}

\begin{table}[t!]
\centering
\scriptsize
\def\arraystretch{0.8}
\begin{tabular}{|@{~}r@{~}|@{~}l@{~}|c|c|c|c|c|}
    \hline
    &                                   & SQL       & CQL       & Mongo     & N1QL      & AQL       \\ \hline
    & \textbf{\texttt{@stored\_named}}  &           &           &           &           &           \\
 1  & \texttt{ .array}                  & \error    & \error    & \error    & \error    & \error    \\
 2  & \texttt{ .bag}                    & \yes      & \yes      & \yes      & \yes      & \yes      \\
 3  & \texttt{ .tuple}                  & \error    & \error    & \error    & \error    & \error    \\
 4  & \texttt{ .scalar}                 & \error    & \error    & \error    & \error    & \error    \\
 5  & \texttt{ .null}                   & \error    & \error    & \error    & \error    & \error    \\
 6  & \texttt{ .missing}                & \error    & \error    & \error    & \error    & \error    \\
\hline \hline
    & \textbf{\texttt{@stored\_bag}}    &           &           &           &           &           \\ \hline
 7  & \texttt{ .element\_array}         & \error    & \error    & \error    & \error    & \partial  \\
 8  & \texttt{ .element\_bag}           & \error    & \error    & \error    & \error    & \partial  \\
 9  & \texttt{ .element\_tuple}         & \yes      & \yes      & \yes      & \yes      & \yes      \\
10  & \texttt{ .element\_scalar}        & \error    & \error    & \error    & \error    & \partial  \\
11  & \texttt{ .element\_null}          & \error    & \error    & \error    & \error    & \partial  \\
12  & \texttt{ .element\_missing}       & \error    & \error    & \error    & \error    & \error    \\
13  & \texttt{ .heterogeneous}          & \error    & \error    & \yes      & \yes      & \yes      \\
\hline \hline
    & \textbf{\texttt{@stored\_array}}  &           &           &           &           &           \\ \hline
14  & \texttt{ .element\_array}         & \error    & \error    & \yes      & \yes      & \yes      \\
15  & \texttt{ .element\_bag}           & \error    & \error    & \error    & \error    & \yes      \\
16  & \texttt{ .element\_tuple}         & \error    & \error    & \yes      & \yes      & \yes      \\
17  & \texttt{ .element\_scalar}        & \error    & \partial  & \yes      & \yes      & \yes      \\
18  & \texttt{ .element\_null}          & \error    & \error    & \yes      & \yes      & \yes      \\
19  & \texttt{ .element\_missing}       & \error    & \error    & \error    & \error    & \error    \\
20  & \texttt{ .heterogeneous}          & \error    & \error    & \yes      & \yes      & \yes      \\
\hline \hline
    & \textbf{\texttt{@stored\_tuple}}  &           &           &           &           &           \\ \hline
21  & \texttt{ .attribute\_array}       & \error    & \yes      & \yes      & \yes      & \yes      \\
22  & \texttt{ .attribute\_bag}         & \error    & \error    & \error    & \error    & \yes      \\
23  & \texttt{ .attribute\_tuple}       & \error    & \error    & \yes      & \yes      & \yes      \\
24  & \texttt{ .attribute\_scalar}      & \yes      & \yes      & \yes      & \yes      & \yes      \\
25  & \texttt{ .attribute\_null}        & \yes      & \yes      & \yes      & \yes      & \yes      \\
26  & \texttt{ .attribute\_missing}     & \error    & \error    & \error    & \error    & \error    \\
\hline
\end{tabular}
\caption{Data Model Feature Matrices}
\label{table:experiments:data-model}
\end{table}

\newcommand{\linedataexp}[1]{%
    \IfEqCase*{#1}{%
    {named}{(lines~1-6)}%
    {bag}{(lines~7-13)}%
    {missing}{(Table~\ref{table:experiments:data-model}, lines~6,12,19,26)}%
    }[\errmessage{Unable to ref #1 for data model experiments}]%
}%

\begin{table}[t!]
\centering
\scriptsize
\def\arraystretch{0.8}
\begin{tabular}{|@{~}r@{~}|@{~}l@{~}|@{~}c@{~}|@{~}c@{~}|@{~}c@{~}|@{~}c@{~}|@{~}c@{~}|}
    \hline
    &                                   & SQL           & CQL           & Mongo         & N1QL          & AQL           \\ \hline
    & \textbf{\texttt{@tuple\_nav}}     &               &               &               &               &               \\
 1  & \texttt{ .support}                & \yes          & \yes          & \yes          & \yes          & \yes          \\
 2  & \texttt{ .absent}                 & \error        & \error        & \missing      & \missing      & \nulloption   \\
 3  & \texttt{ .type\_mismatch}         & \error        & \error        & \inconsistent & \missing      & \error        \\
\hline \hline
    & \textbf{\texttt{@array\_nav}}     &               &               &               &               &               \\
 4  & \texttt{ .support}                & \irrelevant   & \error        & \partial      & \yes          & \yes          \\
 5  & \texttt{ .absent}                 & \irrelevant   & \irrelevant   & \missing      & \missing      & \nulloption   \\
 6  & \texttt{ .type\_mismatch}         & \irrelevant   & \irrelevant   & \missing      & \missing      & \error        \\
\hline \hline
    & \textbf{\texttt{@named\_value}}   &               &               &               &               &               \\
 7  & \texttt{ .restrict\_to\_from}     & \yes          & \yes          & \yes          & \yes          & \yes          \\
\hline \hline
    & \textbf{\texttt{@eq}}             &               &               &               &               &               \\
 8  & \texttt{ .support}                & \yes          & \yes          & \yes          & \yes          & \yes          \\
 9  & \texttt{ .complex}                & \irrelevant   & \error        & \yes          & \yes          & \error        \\
10  & \texttt{ .type\_mismatch}         & \false        & \error        & \false        & \false        & \error        \\
11  & \texttt{ .null\_eq\_value}        & \nulloption   & \false        & \false        & \inconsistent & \nulloption   \\
12  & \texttt{ .null\_eq\_null}         & \nulloption   & \irrelevant   & \true         & \inconsistent & \nulloption   \\
13  & \texttt{ .null\_eq\_missing}      & \irrelevant   & \irrelevant   & \false        & \inconsistent & \irrelevant   \\
14  & \texttt{ .missing\_eq\_value}     & \irrelevant   & \irrelevant   & \false        & \inconsistent & \irrelevant   \\
15  & \texttt{ .missing\_eq\_missing}   & \irrelevant   & \irrelevant   & \true         & \inconsistent & \irrelevant   \\
\hline \hline
    & \textbf{\texttt{@from}}           &               &               &               &               &               \\
16  & \texttt{ .support}                & \yes          & \yes          & \yes          & \yes          & \yes          \\
17  & \texttt{ .named\_value}           & \yes          & \yes          & \yes          & \yes          & \yes          \\
18  & \texttt{ .subquery}               & \yes          & \error        & \error        & \error        & \yes          \\
19  & \texttt{ .collection}             & \yes          & \yes          & \yes          & \yes          & \yes          \\
20  & \texttt{ .at}                     & \yes          & \error        & \error        & \error        & \yes          \\
21  & \texttt{ .bag\_order}             & \counter      & \error        & \error        & \error        & \counter      \\
22  & \texttt{ .coerce\_null\_to\_col}$\ldots$  & \error & \error        & \error       & \error        & \error        \\
23  & \texttt{ .coerce\_missing\_to\_col}$\ldots$ & \irrelevant & \irrelevant & \error  & \error        & \irrelevant   \\
24  & \texttt{ .coerce\_value\_to\_col}$\ldots$ & \error & \error        & \error       & \error        & \error        \\
25  & \texttt{ .tuple}                  & \error        & \error        & \error        & \error        & \error        \\
26  & \texttt{ .coerce\_null\_to\_tuple}& \irrelevant   & \irrelevant   & \irrelevant   & \irrelevant   & \irrelevant   \\
27  & \texttt{ .coerce\_missing\_to\_tuple}& \irrelevant& \irrelevant   & \irrelevant   & \irrelevant   & \irrelevant   \\
28  & \texttt{ .no\_match}              & \nulloption   & \irrelevant   & \irrelevant   & \missing      & \irrelevant   \\
29  & \texttt{ .inner\_correlate}       & \error        & \error        & \error        & \error        & \yes          \\
30  & \texttt{ .left\_correlate}        & \error        & \error        & \error        & \error        & \error        \\
31  & \texttt{ .full\_correlate}        & \error        & \error        & \error        & \error        & \error        \\
32  & \texttt{ .cartesian\_product}     & \yes          & \error        & \error        & \error        & \yes          \\
33  & \texttt{ .inner\_join}            & \yes          & \error        & \error        & \partial      & \error        \\
34  & \texttt{ .left\_join}             & \yes          & \error        & \error        & \partial      & \error        \\
35  & \texttt{ .full\_join}             & \error        & \error        & \error        & \error        & \error        \\
36  & \texttt{ .inner\_flatten}         & \irrelevant   & \error        & \yes          & \yes          & \yes          \\
37  & \texttt{ .outer\_flatten}         & \irrelevant   & \error        & \error        & \error        & \error        \\
\hline \hline
    & \textbf{\texttt{@select}}         &               &               &               &               &               \\
38  & \texttt{ .support}                & \yes          & \yes          & \yes          & \yes          & \yes          \\
39  & \texttt{ .element}                & \error        & \error        & \error        & \yes          & \yes          \\
40  & \texttt{ .attribute}              & \error        & \error        & \error        & \error        & \error        \\
41  & \texttt{ .path}                   & \yes          & \yes          & \yes          & \yes          & \yes          \\
42  & \texttt{ .function\_call}         & \yes          & \partial      & \yes          & \yes          & \yes          \\
43  & \texttt{ .subquery}               & \partial      & \error        & \error        & \partial      & \yes          \\
\hline \hline
    & \textbf{\texttt{@group\_by}}      &               &               &               &               &               \\
44  & \texttt{ .support}                & \yes          & \error        & \yes          & \yes          & \yes          \\
45  & \texttt{ .scalar}                 & \yes          & \irrelevant   & \yes          & \yes          & \yes          \\
46  & \texttt{ .complex}                & \irrelevant   & \irrelevant   & \yes          & \yes          & \yes          \\
47  & \texttt{ .null}                   & \yes          & \irrelevant   & \partial      & \yes          & \yes          \\
48  & \texttt{ .missing}                & \irrelevant   & \irrelevant   & \partial      & \yes          & \irrelevant   \\
49  & \texttt{ .group\_var}             & \error        & \irrelevant   & \yes          & \yes          & \yes          \\
\hline
\end{tabular}
\caption{Query Language Feature Matrices}
\label{table:experiments:query}
\end{table}

\newcommand{\linequeryexp}[1]{%
    \IfEqCase*{#1}{%
    {tuple nav array nav}{(lines~1-6)}%
    {at bag order}{(lines~20-21)}%
    {null eq missing}{(line~13)}%
    }[\errmessage{Unable to ref #1 for query language experiments}]%
}%

To evaluate SQL++'s effectiveness as a configurable and unifying language, we utilize SQL++ config options as a formal expressiveness benchmark of five database:

\renewcommand{\theenumi}{\Alph{enumi}}
\begin{compact_enum}
\item \textbf{SQL} as implemented in MySQL 5.6
\item \textbf{CQL} 3.2.0 of Apache Cassandra 2.1.2
\item the Aggregation Pipeline API of \textbf{MongoDB} 2.6.5
\item \textbf{N1QL} DP4 of Couchbase Server 3.0.1
\item \textbf{AQL} of AsterixDB 0.8.6.
\end{compact_enum}
\renewcommand{\theenumi}{\arabic{enumi}}

All feature matrices have been empirically validated as follows. Classifying a feature as supported (\mbox{\yes}) requires exemplar queries to return results consistent with both SQL++ semantics and the database's documentation. Non-support of a feature (\mbox{\error}) is substantiated by (i) explicit documentation of the restriction (ii) documentation that illustrates the lack of a feature (e.g. the absence of arrays among supported data types), and/or (iii) a query that throws an error. A feature that is \textit{partially supported} (\mbox{\partial}) requires respective substantiation for both supported and unsupported parts. The full benchmark comprising sample data, queries and results, as well as links to supporting documentation and bug reports, is available at \url{http://forward.ucsd.edu/sqlpp}.

Table~\ref{table:experiments:data-model} presents feature matrices that classify the data model of the surveyed databases. We focus on named values that are stored within databases for two main reasons: (i) Since stored values are input by queries, they prominently affect whether query features interact with complex values, heterogeneity etc. (ii) Stored values are independent of query features (subsequently classified in Table~\ref{table:experiments:query}), unlike the values of named views and query results. Parameters in \texttt{@stored\_named} \linedataexp{named} classify whether a stored named value can be an array, bag, tuple, scalar, \texttt{null} or \texttt{missing}. For all stored bags, independenly of whether they are named or nested, \texttt{@stored\_bag} \linedataexp{bag} classifies whether each type of element, as well as heterogeneous elements, are supported. \texttt{@stored\_array} and \texttt{@stored\_tuple} are analogously defined. Collectively, the parameters classify the values that a stored named value can recursively contain, wherein unsupported features indicate a lack of composability.

\highlight{\texttt{@stored\_named}} Following SQL, all databases restrict stored named values to bags (i.e. a SQL table).

\highlight{\texttt{@stored\_bag}} CQL, MongoDB and N1QL follow SQL in restricting bags to only comprise tuples. AQL is classified with partial support, as the restriction applies to named bags but not nested bags. Unlike SQL and CQL, bags can be heterogeneous in MongoDB, N1QL and AQL. 

\highlight{\texttt{@stored\_array}} SQL does not support arrays. A CQL array supports up to 65k scalar elements of the same type (e.g. string). MongoDB/N1QL arrays are heterogeneous and contain arrays, tuples, scalars and \texttt{null}s. An AQL array further supports bag elements.

\highlight{\texttt{@stored\_tuple}} Tuple attributes can be scalars and \texttt{null}s in SQL. In addition, CQL supports arrays, MongoDB and N1QL also support tuples, and AQL further supports bags.

Table~\ref{table:experiments:query} classify query language features, and show how databases are more restrictive than SQL in some ways, but extend beyond SQL in others. We illustrate the classification convention using \texttt{@tuple\_nav} and \texttt{@array\_nav} \linequeryexp{tuple nav array nav}. Parameter \texttt{support} classifies whether tuple/array navigation is supported (\mbox{\yes}), unsupported (\mbox{\error}) or partially supported (\mbox{\partial}). Parameters \texttt{absent} and \texttt{type\_mismatch} are as defined in Figures~\ref{figure:paths:bnf} and \ref{figure:paths:semantics}. For brevity, the options of Figure~\ref{figure:paths:bnf} are abbreviated: \texttt{null} (\texttt{n}), \texttt{missing} (\texttt{m}) and \texttt{error} (\mbox{\error}). In cases where a feature's semantics vary under different circumstances, it is classified as \textit{inconsistent} (\mbox{\inconsistent}). Finally, a feature is \textit{irrelevant} (\mbox{\irrelevant}) if it depends on other unsupported features. For example, since arrays are unsupported in SQL, array navigation is irrelevant.

\highlight{\texttt{@tuple\_nav}} All databases support tuple navigation, but behave differently for absent attributes or navigation from non-tuples. MongoDB has inconsistent semantics: navigating from scalars return \texttt{missing}, yet navigating from arrays return other arrays. Notably, even though \texttt{missing} cannot be stored in MongoDB and N1QL \linedataexp{missing}, it can be an intermediate result during query processing.

\highlight{\texttt{@array\_nav}} CQL supports arrays, but not array navigation. MongoDB only supports array navigation in the \texttt{WHERE} clause, in which an expression language different from that of other clauses has been implemented. In both N1QL and AQL, on the contrary, array navigation is fully supported and behaves symmetrically with tuple navigation.

\highlight{\texttt{@named\_value}} All databases follow SQL such that a named value (which is always a bag) is restricted to being ranged over within the \texttt{FROM} clause, instead of being usable in arbitrary expressions (such as \texttt{table1 = table2}).

\highlight{\texttt{@eq}} The parameters are defined in Figures~\ref{figure:equality:bnf} and \ref{figure:equality:semantics}. In SQL, (i) complex equality is irrelevant due to the named value restriction above (ii) \texttt{1=null}$\rightarrow$\texttt{null} (iii) \texttt{null=null}$\rightarrow$\texttt{null} (iv) \texttt{1='a'}$\rightarrow$\texttt{false}, a peculiarity of MySQL unlike other SQL databases. All other features are irrelevant as complex equality and \texttt{missing} are unsupported. In CQL, \texttt{1=null}$\rightarrow$\texttt{false} (unlike SQL), and \texttt{null=null} is precluded by other limitations (details in the extended version). In MongoDB, equality is cleanly supported through identity comparisons of \texttt{null}/\texttt{missing}. In N1QL, equality on \texttt{null}/\texttt{missing} is broadly inconsistent, as N1QL violates the equivalence $x$\texttt{=}$y$ $\Leftrightarrow$ \texttt{[}$x$\texttt{]=[}$y$\texttt{]}. For example, \texttt{null\_eq\_scalar} is inconsistent because \texttt{1=null}$\rightarrow$\texttt{null} (like SQL), whereas \texttt{[1]=[null]}$\rightarrow$\texttt{false}. Finally, AQL follows SQL equality. Notice that four-valued logic parameters are irrelevant in the 5 databases and have thus been omitted: either complex equality is unsupported, or it is supported but shallow equality does not return \texttt{null}/\texttt{missing} (see N1QL inconsistency above).

\highlight{\texttt{@from}} The parameters classify \texttt{FROM} clause features (Section~\ref{section:from}), and options for \texttt{bag\_order}, \texttt{no\_match}, \texttt{coerce\_null\_to\_collection} etc. are as specified in Figure~\ref{figure:from:bnf}. SQL supports ranging over a named collection, over a subquery, Cartesian product, \texttt{INNER JOIN}, and \texttt{LEFT JOIN}\footnote{\texttt{FULL JOIN} is anomalously unsupported in MySQL.}. MySQL also supports a counter for the iteration order over a bag \linequeryexp{at bag order}. CQL, MongoDB and N1QL are less composable. CQL only supports ranging over a named collection. Likewise for MongoDB, but with the addition of \texttt{INNER FLATTEN}. N1QL further supports joins, but, unlike SQL, restricts them to between primary/foreign keys, and uses \texttt{missing}s for no-matches. Notably, AQL exceeds SQL by supporting the more general \texttt{INNER CORRELATE}. However, AQL does not support \texttt{LEFT}/\texttt{FULL JOIN}. No database supports yet coercions, ranging over tuple attributes, \texttt{LEFT CORRELATE}, or \texttt{OUTER FLATTEN}.

\highlight{\texttt{@select}} SQL and CQL can neither output nor store heterogeneous values. Conversely, MongoDB, N1QL and AQL can both output and store heterogeneous values. All databases follow SQL in supporting paths and functions within \texttt{SELECT}. However, many CQL functions (e.g. \texttt{=}, \texttt{AND}) are supported only in \texttt{WHERE} but not \texttt{SELECT}. Only N1QL and AQL support \texttt{SELECT ELEMENT} to output a collection of non-tuples, and only AQL fully supports subqueries to output nested collections. A SQL subquery must output scalars, and a N1QL subquery must be correlated with the outer query by primary/foreign keys. No database supports yet \texttt{SELECT ATTRIBUTE} to output a tuple of variable-width.

\highlight{\texttt{@group\_by}} CQL does not support \texttt{GROUP BY}. All other databases support grouping all values in their data models. MongoDB, N1QL and AQL also extend beyond SQL to support the \texttt{group} variable. Counterintuitively, MongoDB groups \texttt{null} and \texttt{missing} together as if they were identical, which is contrary to \texttt{@eq.null\_eq\_missing=false} \linequeryexp{null eq missing}.

\balance

\section{Conclusions and Future Work}
\label{section:conclusions}
We described the \texttt{FROM}, \texttt{WHERE}, \texttt{GROUP BY} and \texttt{SELECT} of the configurable, unifying and semi-structured SQL++ query language. SQL++ has picked salient features (composability, semantics without need for schema) from prior query languages - most notably XQuery and OQL. Unlike XQuery's data model, JSON is simple and easily adjusted to be backwards-compatible with SQL, as shown. This enables the SQL++ semantics to be significantly shorter than XQuery, while also offering additional functionalities such as a generalization of SQL's (left) outerjoin and a generalization of SQL's grouping. Furthermore, SQL++ avoids unnecessary extensions over SQL. Rather, it achieves many of its additional capabilities simply by removing semantic restrictions of SQL.

We also showed how appropriate settings of the configuration options morph SQL++ into SQL or any of four NoSQL query languages. The extended version \cite{sqlpp-extended-corr-2015} provides the syntax and semantics of \texttt{ORDER BY}, \texttt{LIMIT}, set operators (\texttt{UNION} etc) and schemas. Furthermore, it provides a SQL++ algebra that (in the spirit of SQL++) is highly compatible with the set-at-a-time processing that SQL engines (based on SQL algebras) use. An earlier extended version \cite{sqlpp-survey-2014} provides the configuration options for an additional six semi-structured databases.

While the NoSQL space is bound to change, the configuration options technique (including the proposed macro options) provides a standing contribution in (a) itemizing and rationalizing the options of a language designer and (b) describing the capabilities and semantics of NoSQL databases, while abstracting away superficial syntactic differences.

 The extended version \cite{sqlpp-extended-corr-2015} will be expanded to describe (using the configuration options technique) the capabilities of selected SQL databases that have recently introduced JSON columns (e.g., \cite{oracle-flexible-schema-cidr-2015}) and respective UDFs. 
Instead of forcing users to learn multiple new primitives (via UDF's) for ranging over various patterns of JSON data and/or constructing various patterns of JSON data, SQL++ lets \texttt{FROM} variables to freely range over any type of data: attribute/value pairs, array elements and index positions in either correlated or uncorrelated fashion, and based on this freedom delivers high expressive power. Similarly, SQL++ allows the \texttt{SELECT} clause to seamlessly create heterogenous and semi-structured JSON by the mere use of nested queries.%

\noindent {\bf Acknowledgements} We thank: Michalis Petropoulos (Pivotal), Yupeng Fu (Palantir) and Romain Vernoux (INRIA) for early contributions to the design and implementation of FORWARD's SQL++ middleware processor; Ashna Jain (UCSD) and Romain Vernoux for their contributions on the experiments. Aditya Avinash for his contributions on the SQL++ reference implementation. Gerald Sangudi (Couchbase), Mike Carey (UCI) and Yannis Katsis (UCSD) for numerous comments and suggestions; Yingyi Bu (UCI) and Jules Testard (UCSD) for the ASTERIX SQL++ implementation. Jacques Nadeau (MapR) for pointing us to use cases requiring iterating over attribute/value pairs; (again) Gerald Sangudi for his work on formally reducing N1QL  to SQL++; Arun Patnaik and Pritesh Maker (Informatica) for discussions on SQL++ as a unifying middleware query language. Finally, Informatica and NSF for their financial support of this work. 

\end{sloppypar}

\bibliographystyle{abbrv}
\bibliography{main}

\end{document}